\documentclass[prd,amsmath,amssymb,nofootinbib,superscriptaddress]{revtex4}
\usepackage{array,mathtools,amssymb,booktabs,multirow,diagbox,hhline,natbib}
\pdfoutput=1

\usepackage{graphicx,graphics,color}% Include figure files
\usepackage{dcolumn}% Align table columns on decimal pointM_\pi
\usepackage{bm}% bold math
\usepackage{bbold}
\usepackage{amsmath}
\usepackage{xcolor}

\allowdisplaybreaks

\newcommand{\Od}{{\cal O}}
\newcommand{\tr}{\mbox{Tr}}

\newcommand{\re}{\mbox{Re}\,}
\newcommand{\mean}[1]{\left\langle{#1}\right\rangle}
\newcommand{\condl}{\mean{\bar q q}}

\newcommand{\condpiquark}{\mean{i\bar q \gamma_5 \tau_1 q}}

\newcommand{\ID}{{\mathbb{1}}}

\newcommand{\gsim}{\raise.3ex\hbox{$>$\kern-.75em\lower1ex\hbox{$\sim$}}}

\begin{document}
%\baselineskip=20pt
% declarations for front matter

\title{On the effective lagrangian at nonzero isospin chemical potential}
\author{A. G\'omez Nicola}
\email{gomez@ucm.es}
\affiliation{Universidad Complutense de Madrid, Facultad de Ciencias F\'isicas, Departamento de F\'isica Te\'orica and
IPARCOS. Plaza de las Ciencias 1, 28040 Madrid, Spain}
\author{A. Vioque-Rodr\'iguez}
\email{avioque@ucm.es}
\affiliation{Universidad Complutense de Madrid, Facultad de Ciencias F\'isicas, Departamento de F\'isica Te\'orica and
IPARCOS. Plaza de las Ciencias 1, 28040 Madrid, Spain}

\begin{abstract}
  We revisit the most general effective lagrangian within Chiral Perturbation Theory at nonzero isospin chemical potential. In addition to the contributions already considered in the literature, we discuss the effects of new terms  allowed by the symmetries,  derived within the external source method including spurion fields, as well as of linear-field corrections. We study the influence of those new contributions on the energy density at zero temperature and observables derived from it, such as the pion and quark condensates and the isospin density. Corrections are shown to be compatible with lattice results, which favor a nonzero value for the  only undetermined  low-energy constant (LEC) to leading order $\Od(p^2)$, rendering in particular a shift of the critical value for Bose-Einstein condensation. To $\Od(p^4)$ we study the physical constraints on the new LEC, which renormalize the energy density and whose numerical effect is estimated within natural values.  The new $\Od(p^4)$ corrections give rise to more significant deviations than those previously considered and remain compatible with  lattice results. 
   \end{abstract}
 \maketitle

\section{Introduction}

The study of the QCD phase diagram has experienced a notable boost over the last decade,  due to the advance of both lattice field theory  at nonzero temperature and chemical potentials \cite{Aoki:2009sc,Borsanyi:2010bp,Bazavov:2014pvz,Bazavov:2018mes,Ratti:2018ksb,Bazavov:2019lgz} 
and  experimental data  of Relativistic Heavy-Ion Collisions (RHIC) within the so called Beam Energy Scan (BES) program probing the transition at chemical freeze-out   \cite{Adamczyk:2017iwn,Andronic:2017pug}.  

Regarding chemical potentials, the main interest has been focused  on the baryon chemical potential $\mu_B$, motivated mostly by the possible existence of a critical point separating the $\mu_B=0$ crossover transition from a first-order one \cite{Ratti:2018ksb,Bazavov:2019lgz}. However, the difficulties for the lattice analyses at $\mu_B\neq 0$ associated to the sign problem \cite{DElia:2002tig,deForcrand:2006pv,Fodor:2004nz,Aarts:2008wh,Bazavov:2018mes} have motivated the study of QCD when chemical potentials associated to additional, physically relevant, charges are present,   which do not present such sign problem. This is the case of isospin $\mu_I$ and strangeness $\mu_S$ chemical potentials, which are actually relevant phenomenological quantities at RHIC at chemical freeze-out, related to electric charge and strangeness conservation \cite{Andronic:2017pug,Bazavov:2014xya}.  Another example is the chemical potential $\mu_5$ associated to chiral imbalance, which has been explored recently both in the lattice \cite{Yamamoto:2011ks,Braguta:2015owi,Astrakhantsev:2019wnp} and within effective theories \cite{Andrianov:2012dj,Espriu:2020dge}. The latter has been mainly motivated by the possible existence of local $P$-breaking regions within RHIC and related phenomena such as the Chiral Magnetic Effect \cite{Kharzeev:2013ffa}.

The case of isospin chemical potential $\mu_I$ has become particularly interesting. As it was first showed in \cite{Son:2000xc} within a leading-order (LO) Chiral Perturbation Theory (ChPT) effective theory approach, QCD at increasing $\mu_I$ exhibits a second-order phase transition from the normal vacuum phase  to a Bose-Einstein Condensation (BEC)  phase for the charged pion modes. The transition point $\mu_c$ predicted by ChPT analyses is at the physical pion mass,  including  next-to-leading order (NLO)  corrections \cite{Adhikari:2019mdk,Adhikari:2020ufo}.  For high enough $\mu_I$, the system would enter a BCS-like phase \cite{Son:2000xc} where effective theory approaches based on the low-density regime, such as ChPT,  are expected to break down \cite{Detmold:2012wc,Carignano:2016lxe}. 

Other recent analyses at $\mu_I\neq 0$ have included finite temperature corrections, allowing to study the effects on the QCD transition \cite{Loewe:2002tw,Fraga:2008be,Cohen:2015soa,Adhikari:2020kdn,Adhikari:2020ufo}. Temperature effects are expected to smooth out the BEC transition, increasing the value of the critical  $\mu_I$ value for BEC  and setting an upper temperature limit for which the BEC phase does no longer take place. In addition, the QCD transition temperature $T_c$ is expected to drop with increasing $\mu_I$.  Other effects considered in this context have been    $\mu_B\neq 0$    \cite{Splittorff:2000mm}  and $\mu_S\neq 0$  \cite{Kogut:2001id,Adhikari:2019mlf} corrections.  A recent analysis  within the NJL model confirms the ChPT results, showing some deviations for $\mu_I\gsim 2M_\pi$ \cite{Lu:2019diy}.

Lattice works have confirmed most of the previous findings. In particular, in \cite{Detmold:2012wc}  the energy density as a function of the isospin density $n_I$, as well as $n_I$ versus $\mu_I$ were investigated at a small but nonzero temperature of around $T\sim$ 20 MeV. Significant deviations  from the  ChPT prediction take place for $\mu_I\gsim 2M_\pi$ and the ratio of the energy density with the Stefan-Boltzmann energy density corresponding to free quarks exhibits a maximum around $\mu_I\simeq 1.3$ $M_\pi$, which was interpreted in \cite{Detmold:2012wc} as the onset of the BEC phase.  Such peak behaviour is well reproduced both within the ChPT and NJL approaches \cite{Lu:2019diy}.  The apparent plateau of that ratio above $\mu_I=3M_\pi$ might indicate a BEC-BCS crossover. 

On the other hand, lattice simulations at finite temperature and nonzero isospin chemical potential confirm the decreasing behaviour of $T_c$ with $\mu_I$ as well as the increase of the BEC onset with temperature 
\cite{Kogut:2004zg,deForcrand:2007uz}. Recent analyses provide results for the pion and quark condensates from $T=113$ MeV up to the QCD transition temperature, confirming  those trends for the $T-\mu_I$ phase diagram \cite{Brandt:2016zdy,Brandt:2017oyy}.  Lattice results for $n_I$ as a function of $\mu_I$ for low temperatures are provided in \cite{Brandt:2018bwq}.   Recent comparisons between lattice data and the ChPT and NJL model approaches can be found respectively in \cite{Adhikari:2019zaj} and \cite{Lopes:2021tro}.

In this work we will discuss some new relevant aspects related to the formulation of the effective ChPT lagrangian  at nonzero isospin density. The usual approach to construct the effective lagrangian follows from considering the isospin contribution in the QCD lagrangian as an external constant vector source  in the $\tau_3$ direction and using the well-know external source method \cite{Gasser:1983yg,Gasser:1984gg} so that the chemical potential enters through covariant derivatives rendering the theory  locally invariant under the chiral symmetry. However, since the $\mu_I$ term in QCD is not invariant under isospin rotations in the $\tau_{1,2}$ directions, the most general effective lagrangian may include additional terms with such symmetry-breaking patttern, which would be multiplied by additional low-energy constants (LEC). A systematic procedure to account for all the possible terms of such type is provided by the so-called spurion method, which is an extension of the external source method 
 developed originally to include correctly the effects of the electromagnetic field in the chiral lagrangian \cite{Ecker:1988te,Urech:1994hd,Meissner:1997fa,Knecht:1997jw}. Following similar ideas, the most general chiral lagrangian for $\mu_5\neq 0$ has been derived up to $O(p^4)$ in \cite{Espriu:2020dge}.

Thus,  the plan of the paper is as follows. In section \ref{sec:lag} we will revisit the construction of the most general $\mu_I\neq 0$ effective lagrangian up to $\Od(p^4)$ including possible new terms of the type commented above, which include both LO and NLO contributions, as well as a new correction coming from the linear terms in the effective lagrangian, which arise to NLO. The effect on different observables steming from the energy density and available in the lattice will be carefully examined to LO in section \ref{sec:lo}, where we will perform a fit analysis to lattice results using the only undetermined LEC to that order.  NLO corrections will be discussed in section \ref{sec:nlo}, including the effect of new LEC associated to additional terms allowed by the symmetries, as well as the  linear correction.  In this paper we will work for $SU(2)$ at $T=0$ to obtain a first glance of those new effects, leaving the finite temperature and strangeness corrections for future work.

\section{Chiral lagrangian  including explicit isospin-breaking operators}
\label{sec:lag}

We start from the QCD lagrangian including a nonzero isospin chemical potential, corresponding to the grand-canonical partition function. The fermionic part of the lagrangian reads 

\begin{equation}
{\cal L}_{QCD}  = \bar q \left(  i\not \! \! D -{\cal M} \right) q+\frac{\mu_I }{2}  \bar q \gamma_0 \tau_3 q
\label{lagmui}
\end{equation}
with $q^T=(u,d)$, $\not \! \! D$ is the covariant derivative corresponding to the gluon field, ${\cal M}$ is the quark mass matrix, which here we take as ${\cal M}=m \ID$ with $m=m_u=m_d$,  and $\tau_k$ are the Pauli matrices. 

The $\mu_I=0$ lagrangian is chiral invariant $SU(2)_L\times SU(2)_R$ for ${\cal M}=0$ (chiral limit) which reduces to the isospin symmetry $SU(2)_{V}$ with $V=L=R$ for nonzero quark masses.  However, for nonzero $\mu_I$, the latter symmetry breaks down to $U(1)_{I_3}$ since the $\mu_I$ term is only invariant under vector transformations in the $\tau_3$ direction. Note also that Lorentz invariance is also broken by the inclusion of the $\mu_I$ term, as a consequence of the preferred reference frame  of the thermal bath at rest.  In addition the isospin chemical potential term  breaks  $C$ invariance, since it is  essentially a charge operator. 

Therefore, the low-energy effective lagrangian must share the above symmetry requirements in the most general way. That is, at a given order in the chiral (low-energy) power counting, one must include all possible operators compatible with the symmetries and their breaking. This can be ensured by following the external source method, where the scalar, pseudoscalar, vector and axial vector sources are  space-time dependent to ensure  local chiral  invariance, as well as Lorentz, $P$, $C$ invariance, of the QCD lagrangian. The effective lagrangian is then constructed out of the most general set of operators satisfying such invariance at a given order in the generic momentum scale $p$, which accounts for meson masses and derivatives, as well as $\mu_I=\Od(p)$ which limits the ChPT analysis to low and moderate values of $\mu_I$, as discussed below. 

An important comment is in order here. The building blocks for constructing such operators are in principle the Goldstone Boson (GB) field operator codified in a $SU(2)$ matrix field $U$, its covariant derivative $d_\mu U$ including the external sources, as well as the flavor matrices entering those external sources, which in the case given by \eqref{lagmui} are the mass matrix ${\cal M}$ and $\frac{\mu_I }{2} \tau_3$.  The latter point, i.e, the fact that one can have additional operators including $\frac{\mu_I }{2} \tau_3$ in a compatible way with the symmetry pattern, is actually one of the main novelties of our present work. One can easily understand this by considering all possible operators to orden $\Od(p^2)$. In addition to the standard terms

\begin{equation}
\tr \left[ d_\mu U d^\mu U^\dagger \right], \tr \left[{\cal M} (U+U^\dagger) \right],
\label{examplesord}
\end{equation}
the following independent term

\begin{equation}
\tr \left[ U \tau_3 U^\dagger \tau_3 \right]
\label{examplenew}
\end{equation}
is also allowed since it breaks  chiral symmetry but preserves $U(1)_{I_3}$ (see details below). The philosophy to include such additional terms is the same as when introducing  electromagnetic corrections to the chiral lagrangian  \cite{Ecker:1988te} and the systematic procedure to account for all those terms consists in introducing the so-called ``spurion" fields $Q_{L,R} (x)$ as additional space-dependent external sources transforming suitably under chiral transformations  \cite{Ecker:1988te,Urech:1994hd,Meissner:1997fa,Knecht:1997jw,Espriu:2020dge}. 

Let us explain this procedure in more detail for our present case. For that purpose, we cast the lagrangian \eqref{lagmui} in terms of external sources as

\begin{equation}
{\cal L}_{QCD}  [v_\mu,A_\mu Q,s] = i\bar q  \not \! \! D   q  + \bar q \left\{ \left[v_\mu (x)+A_\mu (x) Q(x) \right] \gamma^\mu - s(x)+i\gamma_5 p(x)\right\}q
\label{sourcelag}
\end{equation} 
where $v_\mu\in SU(2)$. 
The above lagrangian corresponds to the choice relevant for this work, $Q_L=Q_R=Q$, and we have set the axial source $a_\mu=0$ with respect to the general source lagrangian considered in \cite{Urech:1994hd,Meissner:1997fa,Knecht:1997jw,Espriu:2020dge}, where we have kept a nonzero pseudoscalar source $p(x)$ in order to derive expectation values of pionic fields, such as the pion condensate. 

The lagrangian in \eqref{lagmui} corresponds to the particular choice $v_\mu+A_\mu Q= \frac{\mu_I }{2} \tau_3$, $s={\cal M}$ and $p=0$. Thus, after the general effective lagrangian is constructed,  we will choose, without loss of generality, 

\begin{equation} 
v_\mu=0, \quad A_\mu=\Lambda \delta_{\mu 0}, \quad Q= \frac{\mu_I}{2\Lambda} \tau_3 
\label{choice}
\end{equation}
consistently with our power counting as  long as $\mu_I \ll \Lambda$. The choice of the parameter $\Lambda$ is irrelevant, as it should, since it will be absorbed in the new independent LEC involved, which will have to be determined. The important point is to include $\mu_I$ in the $Q$ contribution, consistently with the counting $\mu_I=\Od (p)$. In addition, as customary we will write $p(x)=j \tau_1$ so that thaking derivatives with respect to $j$ we reproduce expectation values of the $\pi_1$ field, which is the direction we have chosen for the condensed field. 

The  lagrangian \eqref{sourcelag} can be made  locally invariant under chiral $SU(2)_L\times SU(2)_R$ rotations of the quark and source fields. Here, it is enough to restrict to the vector subgroup $g_L=g_R=g\in SU(2)$, under which the lagrangian in  \eqref{sourcelag} can be made invariant by considering the following transformations

\begin{eqnarray}
q(x)&\rightarrow& g(x) q(x) \nonumber\\
\chi(x)&\rightarrow& g (x) \chi(x) g^\dagger (x) \nonumber \\
v_\mu (x) &\rightarrow& g(x) v_\mu (x) g(x)^\dagger + ig(x) \partial_\mu g(x)^\dagger \nonumber\\
Q (x) &\rightarrow&  g(x) \left[Q(x) \right] g^\dagger (x) 
\label{fieldtrans} 
\end{eqnarray}
with $\chi(x)=2B_0 \left[s(x)+ip(x) \right]$ and $B_0=M^2/(2m)$ with $M$ the tree level pion mass. In addition, one has to consider the $C$ and $P$ transformations of those fields, given in \cite{Urech:1994hd,Meissner:1997fa,Knecht:1997jw,Espriu:2020dge}, leading to a $C, P$ invariant lagrangian. 

The key point here is that the $v_\mu$ and $A_\mu Q$ sources can be transformed independently (we choose here to leave $A_\mu$ invariant for simplicity, which does not affect our arguments). As we are about to see, those ``spurion"  transformations are essential to ensure that all possible operators are accounted for. 

Now, the chiral lagrangian can be constructed order by order in the chiral expansion.  The building blocks and their chiral counting are  the pseudo-GB field $U\in SU(2)$, which is $\Od(1)$, its covariant derivative

\begin{eqnarray}
d_\mu U&=&\partial_\mu U+i[U, v_\mu+ A_\mu Q],
\label{covderu}
\end{eqnarray}
which is $\Od(p)$, and the external fields, which in this case are $\chi=\Od(p^2)$ and $v_\mu=\Od(p)$, $A_\mu Q=\Od(p)$, where we follow the same convention as in \cite{Urech:1994hd,Meissner:1997fa,Knecht:1997jw,Espriu:2020dge} assigning the counting $A_\mu=\Od(1)$, $Q=\Od(p)$ as commented above. In addition, the following covariant derivative of the $Q$ field

\begin{eqnarray}
c_{\mu} Q&=&\partial_\mu Q -i [v_\mu, Q]
\label{covderQ}
\end{eqnarray}
which is $\Od(p^2)$, has to be formally considered, although in practice it will not enter our effective lagrangian at the order considered here.

The chiral transformations of the $U$ field, i.e., $U \rightarrow g_R U g_L^\dagger$, as well as those of the external sources and the covariant derivatives, and their $C$, $P$ transformations,  allow to construct the most general effective lagrangian which is locally chiral and $C,P$ invariant for our present case.  In particular, regarding the external sources, in addition to the usual corrections coming from the covariant derivative \eqref{covderu}, which depends only on $v_\mu+A_\mu Q$, such as the first term in \eqref{examplesord},  additional  terms are allowed, such as \eqref{examplenew}. That term comes from the invariant operator 
$$\tr \left[ Q U Q U^\dagger \right]$$ 
which will be  therefore multiplied by $\mu_I^2$ times an independent LEC.

Let us now proceed order by order within this scheme, following the discussion in \cite{Espriu:2020dge} which considers all possible terms coming from arbitrary constant $Q_{L,R}$ fields.  At the lowest $\Od(p^0)$ order, only trivial $\mu_I$-independent terms can be constructed out of the $U$ fields. At $\Od(p)$ the only allowed structure is $\tr (Q)$, which according to \eqref{choice} would give rise to a term linear in $\mu_I$ in the energy density and therefore would imply a nonzero isospin density for $\mu_I=0$, which is ruled out by the lattice \cite{Detmold:2012wc,Brandt:2018bwq}. Therefore, demanding $n_I(0)=0$ we will fix to zero the LEC multiplying that term and ignore it in what follows.

To $\Od(p^2)$, in addition to the structures \eqref{examplesord} and \eqref{examplenew}, the terms $\tr (Q^2)$ and $(\tr Q)^2$ are also allowed, the latter vanishing with our choice \eqref{choice}.  Therefore, the most general $\Od(p^2)$  lagrangian at nonzero $\mu_I$ is given by

\begin{equation}
{\cal L}_2=\frac{F^2}{4} \tr \left[ d_\mu U d^\mu U^\dagger + \chi^\dagger U+\chi U^\dagger  + \frac{1}{2} a_1 \mu_I^2  U \tau_3 U^\dagger \tau_3 \right] + \frac{1}{4} a_2 F^2 \mu_I^2
\label{L2}
\end{equation}
%\anote{corregido el coeficiente de $a_2$, OK con Andrea}
where  $F$ is the tree-level pion decay constant. $a_1$ and $a_2$ are new low-energy constants to be determined below and we have included a $F^2$ factor in front of the new terms to keep easy track of the  contributions of different chiral orders to the observables, so that the chiral expansion  can be parametrically tracked   in powers of  $F^2$. 

Note also that the equations of motion (eom) derived from the above lagrangian are modified with respect to the $\mu_I=0$ case as

\begin{equation}
(d_\mu d^\mu U^\dagger)U-U^\dagger d_\mu d^\mu U=\chi^\dagger U-U^\dagger \chi+\frac{1}{2}\tr\left[U^\dagger\chi-\chi^\dagger U\right]-\frac{1}{2}a_1 \left(U^\dagger \tau_3 U \tau_3-\tau_3 U^\dagger \tau_3 U\right)
\label{eom} 
\end{equation}

The above $\Od(p^2)$ eom can be used in the construction of the  $\Od(p^4)$ lagrangian to eliminate some operators in favor of a minimal set \cite{Urech:1994hd} together with standard $SU(2)$ operator identities. 

In principle, the following $\Od(p^3)$ operators are also allowed \cite{Espriu:2020dge}:

\begin{align}
\tr \left(\chi^\dagger U+\chi U^\dagger \right)\tr (Q), \quad 
\tr \left(d_\mu U d^\mu U^\dagger \right)\tr (Q), \quad  \tr \left( Q U Q U^\dagger \right) \tr (Q), \quad 
 \tr \left( Q^2  \right) \tr (Q), \quad \left[\tr (Q) \right]^3
\label{op31}\\
\tr \left[ Q \left(d_\mu U d^\mu U^\dagger + d_\mu U^\dagger d^\mu U \right) \right], \quad 
	\tr \left[Q^2\left( UQU^\dagger + U^\dagger Q U \right) \right],  \quad  \tr \left[ Q\left(\chi^\dagger U+\chi U^\dagger + U \chi^\dagger + U^\dagger \chi  \right)\right]
\label{op32}
\end{align}

The five operators in \eqref{op31} vanish trivially with $Q$ in \eqref{choice}. In addition, one can check that the operators in \eqref{op32} also vanish for $d_\mu U$ and $U$ in $SU(2)$ and $Q$ in \eqref{choice} as long as we remain in the isospin limit  $m_u=m_d$, as we will do here consistently with isospin conservation. The last term in \eqref{op32} is actually proportional to $\tr (\tau_3 {\cal M})=m_u-m_d$.

As for the $\Od(p^4)$ lagrangian,  following the derivation in \cite{Espriu:2020dge} we have in our present case

\begin{eqnarray}
{\cal L}_4&=&{\cal L}_4^0+{\cal L}_4^Q \nonumber \\
{\cal L}_4^0&=& \frac{l_1}{4} \left[\tr \left( d_\mu U d^\mu U^\dagger\right)\right]^2 + \frac{l_2}{4} \tr \left( d_\mu U d_\nu U^\dagger\right) \tr \left( d^\mu U d_\nu U^\dagger\right) + \frac{l_3+l_4}{16} \left[\tr\left(\chi^\dagger U+\chi U^\dagger \right)\right]^2 \nonumber\\ 
&+& 
\frac{l_4}{8} \tr \left( d_\mu U d^\mu U^\dagger\right)\tr\left(\chi^\dagger U+\chi U^\dagger \right)
-\frac{l_7}{16} \left[\tr\left(\chi^\dagger U-\chi U^\dagger \right)\right]^2
\nonumber\\ 
&+& 
\frac{h_1+h_3-l_4}{4}\tr\left(\chi^\dagger \chi\right)+\frac{h_1-h_3-l_4}{2}\re\det\chi
\label{L4usual} \\
{\cal L}_4^Q&=& q_1 \Lambda^2 \tr \left( d_\mu U d^\mu U^\dagger\right)\tr\left(Q^2\right) + q_2 \Lambda^2 \tr \left( d_\mu U d^\mu U^\dagger\right) 
\tr\left(   QUQU^\dagger  \right) \nonumber\\
&+& q_3 \Lambda^2 \left[\tr \left( d_\mu U Q U^\dagger\right) \tr \left( d^\mu U Q U^\dagger\right) +  \tr \left( d_\mu U^\dagger Q U\right) \tr \left( d^\mu U^\dagger Q U\right) \right] + q_4 \Lambda^2  \tr \left( d_\mu U^\dagger Q U\right)  \tr \left( d^\mu U Q U^\dagger\right)
\nonumber\\
&+& q_5 \Lambda^2  \tr\left(\chi^\dagger U+\chi U^\dagger \right) \tr\left(Q^2\right) + q_6 \Lambda^2 \tr\left(\chi^\dagger U+\chi U^\dagger \right) \tr\left(   QUQU^\dagger  \right) 
\nonumber\\
&+& q_7 \Lambda^2 \tr\left[\left(\chi^\dagger U-U^\dagger\chi\right)QU^\dagger Q U + \left(\chi U^\dagger-U\chi^\dagger\right)QU Q U^\dagger \right]+q_8\Lambda^4 \left[\tr \left(Q^2\right)\right]^2 +  q_9\Lambda^4  \tr\left(Q U Q U^\dagger \right) \tr \left(Q^2\right) 
\nonumber\\
&+& q_{10} \Lambda^4 \left[ \tr\left(Q U Q U^\dagger \right)\right]^2 
\label{L4Q}
\end{eqnarray}
with $Q$ in \eqref{choice}.   The usual terms considered in the literature are included in the $ {\cal L}_4^0$ lagrangian, for which we have used the basis in \cite{Scherer:2002tk} 
\footnote{In the basis used in \cite{Espriu:2020dge} there is a typo in the LEC multiplying the $\left[\tr\left(\chi^\dagger U-\chi U^\dagger \right)\right]^2$ operator in eq.(4.1) in that paper, which should read $-(l_4+l_7)/16$.}, whereas ${\cal L}_4^Q$ contains all possible operators including explicitly the $Q$ field and whose influence on different observables will be analyzed in detail in section \ref{sec:numres}.   With respect to the basis considered in \cite{Adhikari:2019mdk,Adhikari:2020ufo,Adhikari:2020kdn}, the $h_1$ constant considered in those papers corresponds to $h_1-l_4$ here.

In addition, following also the derivation in \cite{Son:2000xc,Adhikari:2019mdk,Adhikari:2020ufo,Adhikari:2020kdn}, we allow for a nontrivial vacuum configuration of the GB field, which will parametrize the pion condensed phase. Namely, we write

\begin{eqnarray}
U(x)&=&A \exp{\left[i \frac{ \tau^a \pi^a (x)}{F}\right]} A \nonumber \\
A&=& \cos(\alpha/2) \ID + i \sin(\alpha/2) \tau_1
\end{eqnarray}
with $\pi^a$ the pion fields, so that the condensed phase is characterized by a nonzero value of the $\alpha$ angle, which is determined by minimizing the vacuum energy density at a given chiral order, from which  we will actually obtain the main properties of interest here. In the present work we are interested in the vacuum energy density  at zero temperature, i.e., 

\begin{equation}
\epsilon(\mu_I,j)=-\lim_{T\rightarrow 0, V\rightarrow\infty}\frac{T}{V}\log Z (T,\mu_I,j)=\epsilon_2+\epsilon_4+ \cdots 
\label{fed}
\end{equation}
where $Z (T,\mu_5)$ is the euclidean partition function constructed out of the effective lagrangian for nonzero $\mu_I$ and $\epsilon_n=\Od(p^n)$. The above definition of the vacuum energy density is equivalent to the effective potential considered in \cite{Son:2000xc,Adhikari:2019mdk,Adhikari:2020ufo,Adhikari:2020kdn}. 

The observables of interest we can calculate in both phases are the quark and pion condensates and the isospin density, which are  given respectively by

\begin{eqnarray}
\condl (\mu_I)&=&  \langle \bar u u + \bar d d \rangle = \left.\frac{\partial \epsilon(\mu_I,j)}{\partial m}\right.=2B_0 \left.\frac{\partial \epsilon(\mu_I,j)}{\partial M^2}\right.
\label{conddef} \\
\condpiquark (\mu_I)&=&  \left.\frac{\partial \epsilon(\mu_I,j)}{\partial j}\right.
\label{condpidef}\\
n_I(\mu_I)&=&-\left.\frac{\partial \epsilon(\mu_I,j)}{\partial \mu_I}\right.
\label{densitydef}
\end{eqnarray}

%
%\begin{eqnarray}
%U&\rightarrow& g_R U g_L^\dagger\nonumber\\
%%Q_I&\rightarrow& g_IQ_I g_I^\dagger \quad (I=L,R)\nonumber\\
%%\chi&\rightarrow& g_R \chi g_L^\dagger \nonumber \\
%d_\mu U&\rightarrow& g_R d_\mu U g_L^\dagger\nonumber\\
%c_{\mu I} Q_I&\rightarrow& g_I  c_{\mu I}  Q_I g_I^\dagger  \quad (I=L,R) \nonumber\\
%G_{\mu\nu}^I&\rightarrow &g_I G_{\mu\nu}^I g_I^\dagger
%\label{chiraltrans}
%\end{eqnarray}
%

\section{Leading order results}
\label{sec:lo}

The lowest order  of the energy density, $\epsilon_2$, parametrically of $\Od(F^2)$, is given just by minus the constant part of the ${\cal L}_2$ lagrangian in \eqref{L2}:

\begin{equation}
\epsilon_2 (\mu_I,j)=-\frac{F^2}{4} \left[  \mu_I^2 
\left( 1+a_2-(1-a_1)\cos (2\alpha)
\right)
+4M^2\cos\alpha+8B_0 j \sin\alpha
\right]
\label{energylo}
\end{equation}

The value $\alpha_0^{LO}$ minimizing the above expression with respect to $\alpha$ is given for $j=0$ by

\begin{equation}
\cos\alpha_0^{LO}=\left\{
\begin{array}{cl}
1 &\displaystyle \quad \mbox{for} \quad  \mu_I<\mu_c =\frac{M}{\sqrt{1-a_1}}\\
\displaystyle
\frac{M^2}{(1-a_1)\mu_I^2} &\displaystyle  \quad \mbox{for} \quad \mu_I>\mu_c \\
\end{array}
\right.
\label{alpha0} 
\end{equation}

Therefore, at this order, the constant $a_1$ displaces the critical BEC value $\mu_c$ from the squared pion mass.  We will see below that this is perfectly compatible with lattice results, which  will constrain the $a_1$ value and its uncertainty. Note that the upper bound $a_1<1$ arises here from the very existence of a BEC phase. 

The above result is also consistent with the modifications of the leading order pion dispersion relations steming from the $\mu_I$-dependent lagrangian above. In fact, let us consider the linear and quadratic terms in the pion field, from the ${\cal L}_2$ lagrangian in \eqref{L2}:

\begin{eqnarray}
{\cal L}_2^{lin}&=&-F\sin\alpha\left[M^2- (1-a_1)\mu_I^2\cos\alpha\right]\pi_1 (x)+F \mu_I \partial_0\pi_2(x)\sin\alpha+2B_0 F j  \pi_1 (x) \cos\alpha
\label{L2lin}\\
{\cal L}_2^{quad}&=& \frac{1}{2}\partial_\mu\pi^a \partial^\mu \pi_a + \frac{1}{2}m_{12}\left[\pi_1(x)\partial_0\pi_2(x)-\pi_2(x)\partial_0\pi_1(x)\right]-\frac{1}{2}\left[m_1^2\pi_1^2(x)+m_2^2\pi_2^2(x)+m_3^2\pi_3^2(x)\right], 
% \nonumber\\ &-&B_0 j  \pi^a (x)\pi_a (x) \sin\alpha
\label{L2quad}
\end{eqnarray}
where, following the notation in \cite{Adhikari:2019mdk}, we have

\begin{eqnarray}
m_{12}&=&2 \mu_I \cos\alpha \label{m12}\\
m_1^2&=&M^2\cos\alpha -(1-a_1)\mu_I^2\cos(2\alpha)+2B_0 j \sin\alpha
\label{m1sq}\\
m_2^2&=&M^2\cos\alpha -(1-a_1)\mu_I^2\cos^2\alpha+2B_0 j \sin\alpha
\label{m2sq}\\
m_3^2&=&M^2\cos\alpha +(1-a_1)\mu_I^2\sin^2\alpha+2B_0 j \sin\alpha
\label{m3sq}
\end{eqnarray}

Now, we follow the same steps as in \cite{Adhikari:2019mdk} to obtain the pion dispersion relation in terms of the parameters $m_{12},m_{1,2,3}^2$. Note that to leading order, we can replace $\alpha=\alpha_0^{LO}$, which in particular cancels the contribution proportional to $\pi_1(x)$ in  \eqref{L2lin} so that such  linear lagrangian becomes a total derivative in the condensed phase and vanishes in the normal phase. Therefore, at this order, the linear lagrangian does not play any role for the dispersion relation nor in the vacuum energy density. As we will discuss in section \ref{sec:lin}, this does no longer hold to NLO, so that term will have to be included.

We then get for the dispersion relation of charged and neutral pions to leading order, respectively, 

\begin{eqnarray}
E_{\pm}^2(p)&=&p^2+\frac{1}{2}\left(m_1^2+m_2^2+m_{12}^2\right)\pm\frac{1}{2} \sqrt{4p^2m_{12}^2+\left(m_1^2+m_2^2+m_{12}^2\right)^2-4m_1^2m_2^2}
\label{Epm}\\
E_0^2(p)&=&p^2+m_3^2
\label{E0}
\end{eqnarray}
 where $p\equiv \vert \vec{p} \vert$. Now, setting $\alpha=\alpha_0^{LO}$ given in \eqref{alpha0}  and $p=0$ we get then the dependence of the static pion masses on the isospin chemical potential, now including the correction from the $a_1$ term, which for $j=0$ read

\begin{equation}
\left\{ 
\begin{array}{ccl}
$$
M_{\pm}^2&=&M^2 + (1+a_1)\mu_I^2 \pm  2\mu_I\sqrt{M^2+a_1\mu_I^2}\\
\\
M_0^2&=&M^2 
$$
\end{array}
\quad  \mu_I<\mu_c
\right.
\label{massesbelow}
\end{equation}

\begin{equation}
\left\{ 
\begin{array}{ccl}
$$
M_+^2&=& \displaystyle (1-a_1)\mu_I^2+ \frac{3+a_1}{(1-a_1)^2} \frac{M^4}{\mu_I^2} \\
\\
M_-^2&=& 0
\\
\\
M_0^2&=&(1-a_1)\mu_I^2
$$
\end{array}
\quad \mu_I>\mu_c
\right.
\label{massesabove}
\end{equation}

\begin{figure}[h]
 \centerline{ \includegraphics[width=0.45\textwidth]{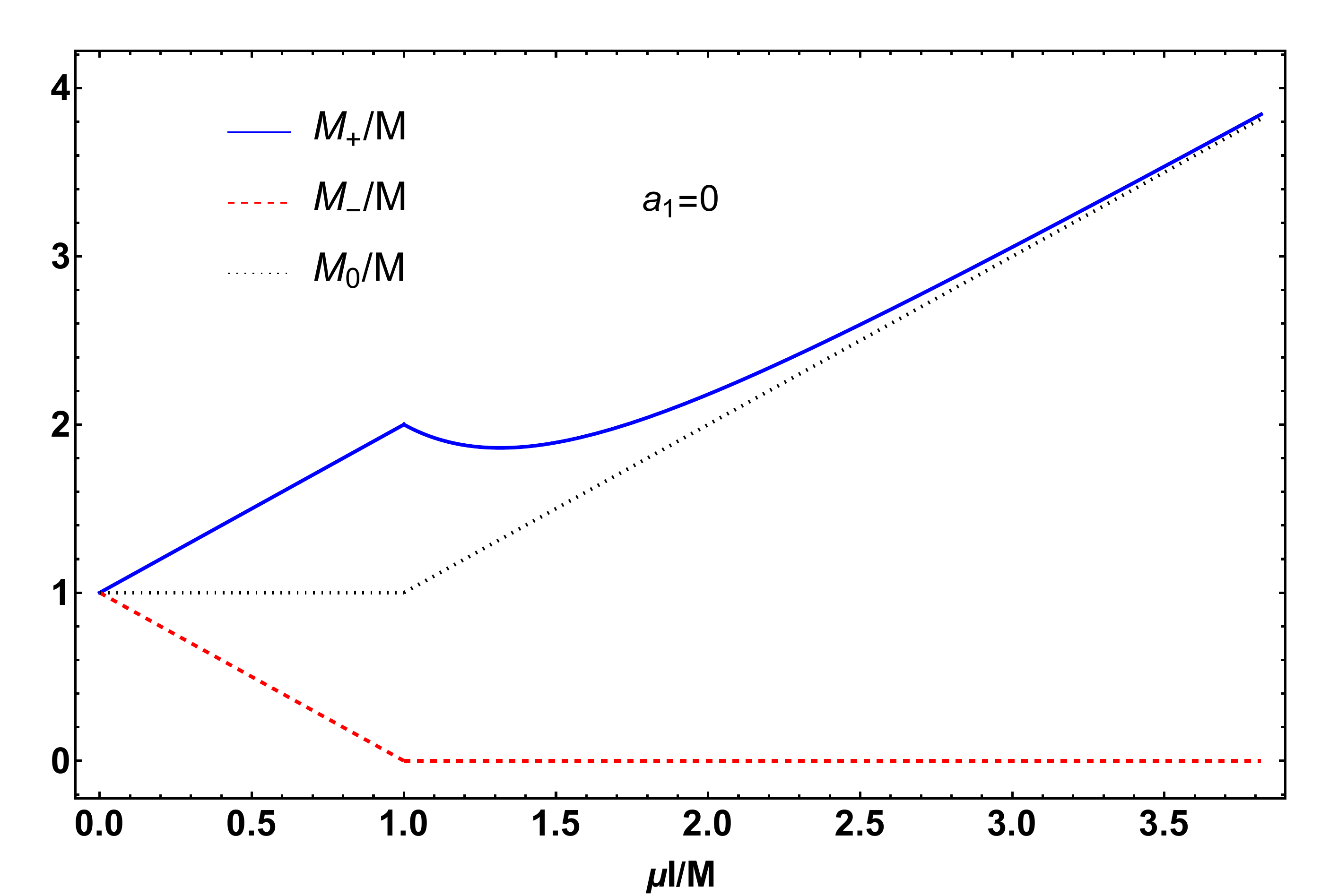}
  \includegraphics[width=0.45\textwidth]{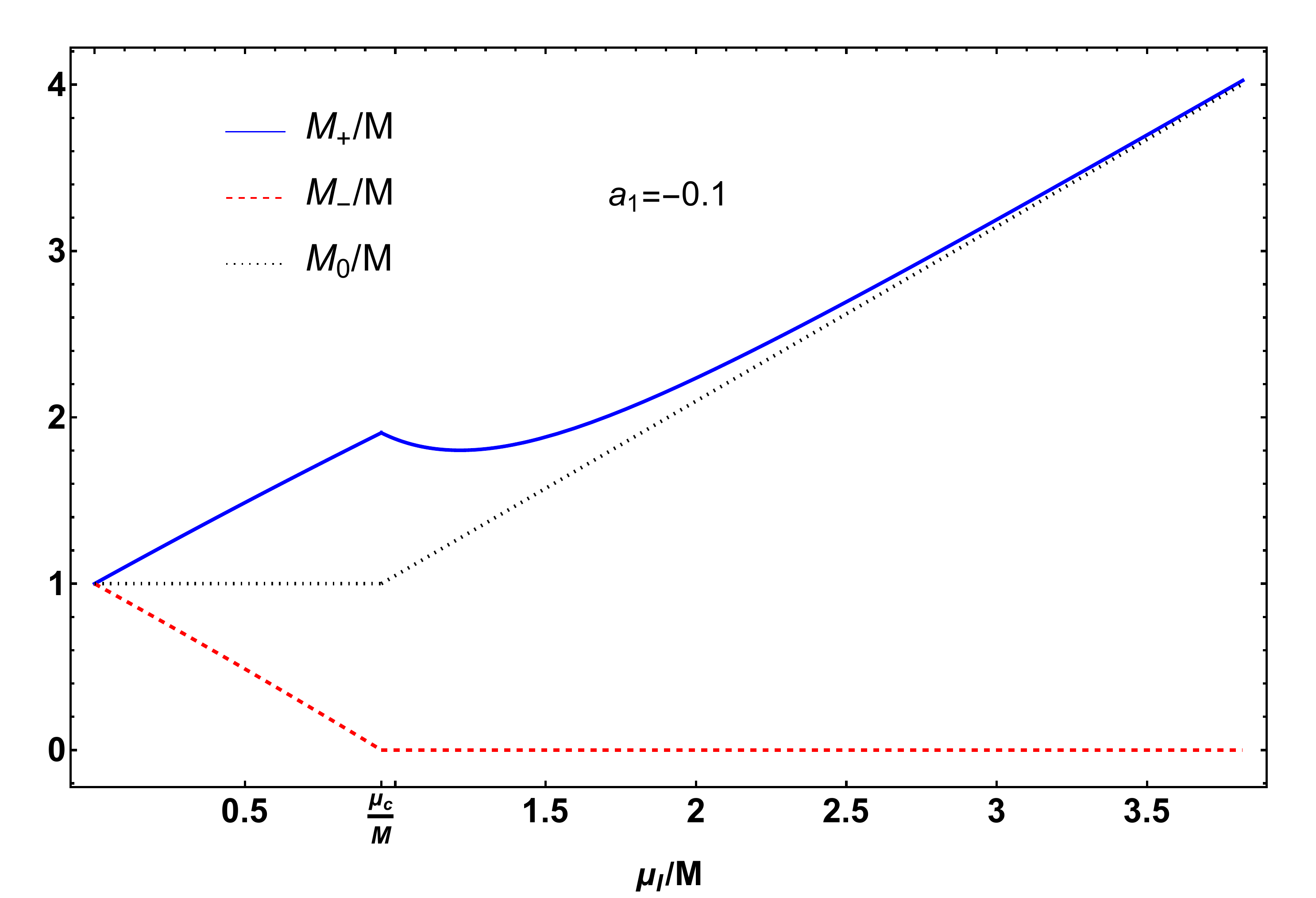}}
  \caption{Dependence of pion masses with isospin chemical potential to leading order in ChPT with and without the extra term, for a sample value $a_1=-0.1$.}
  \label{Figure:masseslo}
\end{figure}

Thus, as in the $a_1=0$ case, the vanishing mass of one of the charged pions signals the onset of BEC condensation, as a Goldstone mode corresponding to the $U(1)_{I_3}$ spontaneous symmetry breaking of the vacuum in that phase, i.e, with a nonzero pion expectation value.  The above mass dependence is plotted in Fig. \ref{Figure:masseslo} for a sample value of $a_1=-0.1$, compared to the $a_1=0$ case,  the qualitative dependence with $\mu_I$ being quite similar, although note that $a_1$ introduces a non-polynomial dependence below $\mu_c$. 

%\anote{ese t\'ermino corregida errata, OK Andrea}

From the energy density  in \eqref{energylo} we get the LO to the quark and pion condensates and the isospin density. Namely, from  \eqref{conddef}, \eqref{condpidef}, \eqref{densitydef}, we get, replacing $\alpha=\alpha_0$ and for $j=0$,

\begin{eqnarray}
\condl^{LO} (\mu_I)&=&  
\left\{
\begin{array}{cl}
-2B_0 F^2 &\displaystyle \quad \mu_I<\mu_c \\ \\
\displaystyle
\frac{-2B_0 F^2M^2}{(1-a_1)\mu_I^2} &\displaystyle   \quad \mu_I>\mu_c \\
\end{array}
\right.
\label{condlo} \\
\condpiquark^{LO} (\mu_I)&=& 
\left\{
\begin{array}{cl}
0 &\displaystyle \quad \mu_I<\mu_c \\ \\
\displaystyle
-2B_0 F^2\sqrt{1-\frac{M^4}{(1-a_1)^2\mu_I^4}}&\displaystyle   \quad \mu_I>\mu_c \\
\end{array}
\right.
\label{condpilo}\\ \nonumber \\
n_I^{LO}(\mu_I)&=&
\left\{
\begin{array}{cl}
\displaystyle
\frac{1}{2}(a_1+a_2) F^2 \mu_I  &\displaystyle \quad \mu_I<\mu_c \\ \\
\displaystyle
\frac{1}{2} F^2 \mu_I\left(2-a_1+a_2-\frac{2M^4}{(1-a_1)\mu_I^4}\right) &\displaystyle   \quad \mu_I>\mu_c \\
\end{array}
\right.
\label{densitylo}
\end{eqnarray}

Note that the only dependence with $a_2$ shows up in the pion density, which should remain zero below the BEC point. Therefore, with that physical requirement we fix

\begin{equation}
a_2=-a_1+\Od(1/F^2)
\label{locons}
\end{equation}
up to higher order corrections, which leaves us only with one free parameter at this order. 

\subsection{LO fits to lattice}
\label{sec:LOfits}

Here, we will discuss more quantitatively the effects of the new terms to leading order, by contrasting the results for different observables with those obtained in the lattice.  For that purpose, we have considered the most recent lattice results. In particular, we will use the lattice points provided in \cite{Adhikari:2020ufo}  at $T=0$, coming from the collaboration \cite{Brandt:2018bwq}, for the quark and pion condensates. The latter are given for a finite pionic source. We take $j=0.00517054 M_\pi$, one of the two values considered in \cite{Adhikari:2020ufo}. As for the isospin density, we compare with the results quoted in  \cite{Brandt:2018bwq} for $j=0$ and $T$ small enough to provide an accurate enough description of the $T=0$ case.  We also fix the numerical values of $M_{\pi}=131 \ \text{MeV}$, $F_{\pi}=90.51 \ \text{MeV}$ and $m=3.47 \ \text{MeV}$ for an easier comparison with the results in \cite{Adhikari:2020ufo,Brandt:2018bwq}. In addition, we assign a characteristic uncertainty of $5\%$ to the lattice data, following again \cite{Adhikari:2020ufo,Brandt:2018bwq}.

As discussed above, to LO the new contributions considered here are parametrized only in the constant $a_1$. The relevant question  whether $a_1\neq 0$ is preferred for lattice results over the standard choice $a_1=0$. Recall that $a_1\neq 0$ implies in particular a displacement in the critical value $\mu_c$ with respect to the pion mass.  Actually, it is worth mentioning here that in some recent lattice analysis \cite{Brandt:2017oyy} the condition $\mu_c=M_\pi$ is actually imposed,  not from a physical requirement but from the ChPT result, i.e., a sufficiently small variation of $\mu_c\neq M_\pi$ could be equally assumed.  In fact, as we are about to see, fits to lattice points favor a negative nonzero value for $a_1$, which according to  \eqref{alpha0} yield $\mu_c<M_\pi$ to LO, consistently for instance with the position of the lowest lattice point in the isospin density given in \cite{Brandt:2018bwq}.

\begin{figure}[h!]
\begin{center}
	%\begin{minipage}{0.48\textwidth}
		\includegraphics[width=0.48\textwidth]{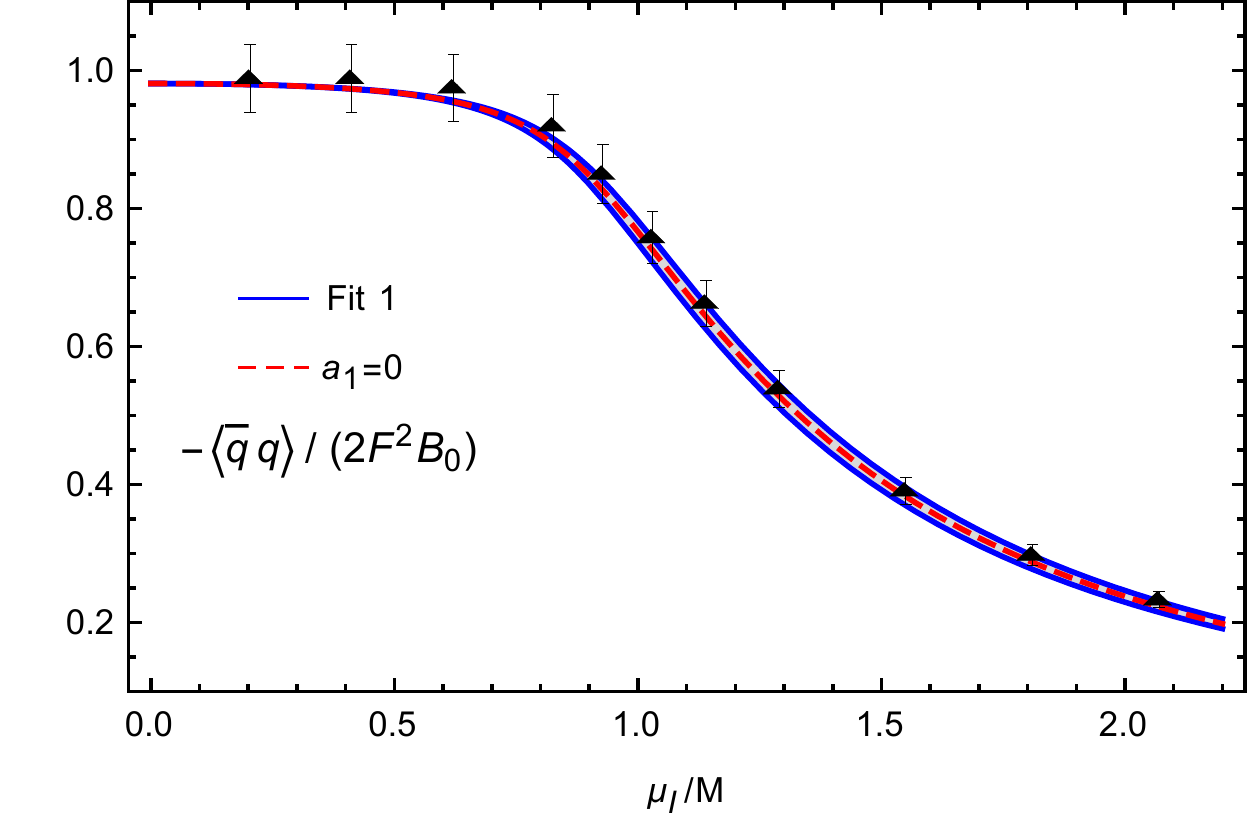}
	% \end{minipage}
	% \begin{minipage}{0.48\textwidth}
		\includegraphics[width=0.48\textwidth]{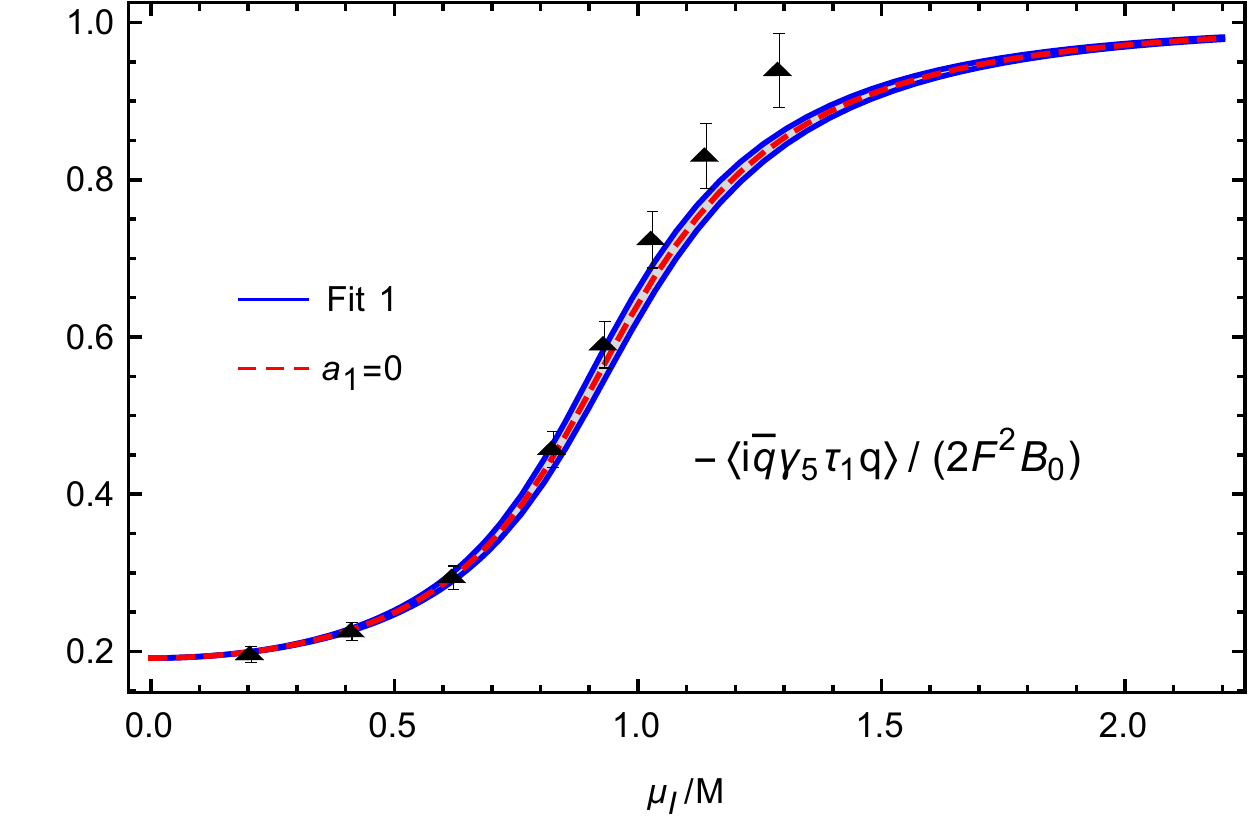}
	%\end{minipage}
	\end{center}
	\caption{Combined fit of the quark condensate (left) and the pion condensate (right) for $j=0.00517054 M_\pi$. Lattice points are taken from \cite{Adhikari:2020ufo}.}
	\label{fig:comfitcondqqpi}
\end{figure}

\begin{figure}[h!]
	\begin{center}
		\includegraphics[width=0.8\textwidth]{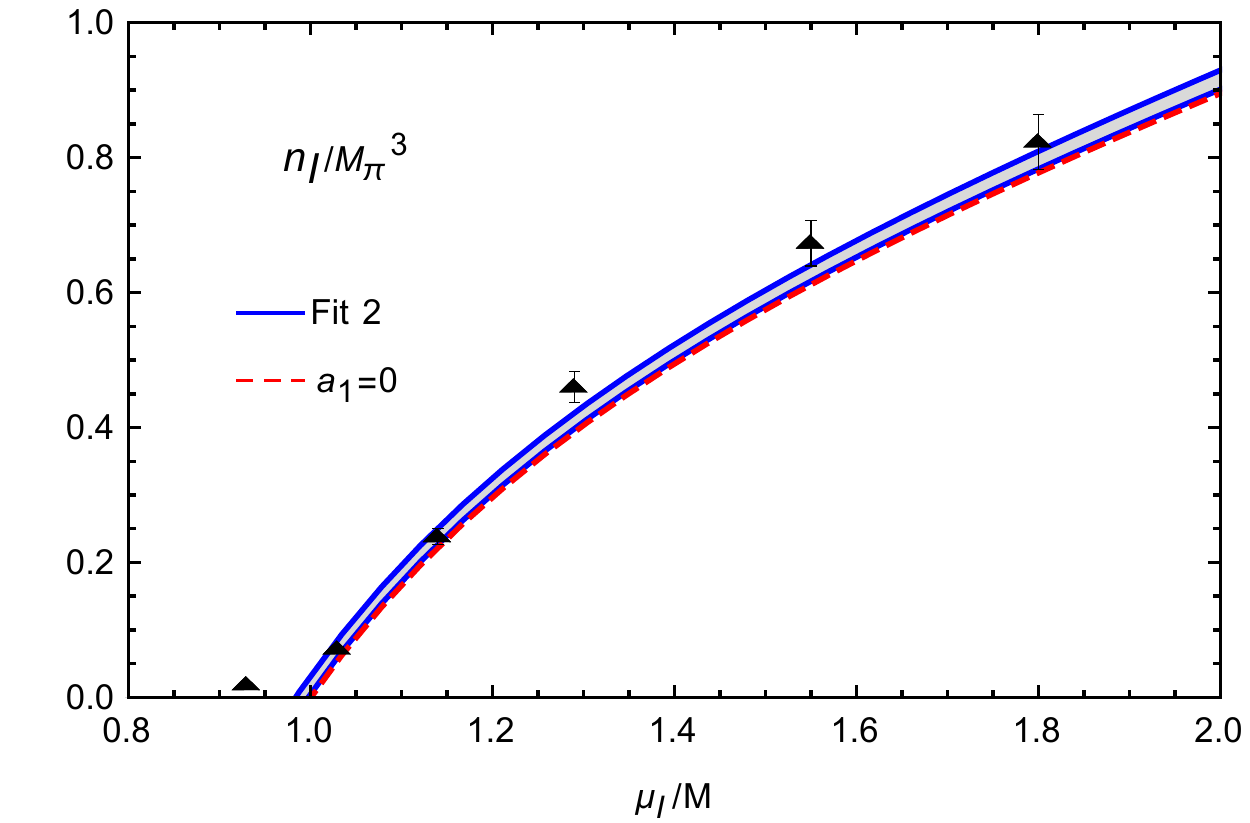}
	\end{center}
	\caption{Fit of the isospin density for $j=0$. Lattice points are taken from \cite{Brandt:2018bwq}.}
	\label{fig:fitnI}
\end{figure}

\begin{figure}[h!]
		\includegraphics[width=0.48\textwidth]{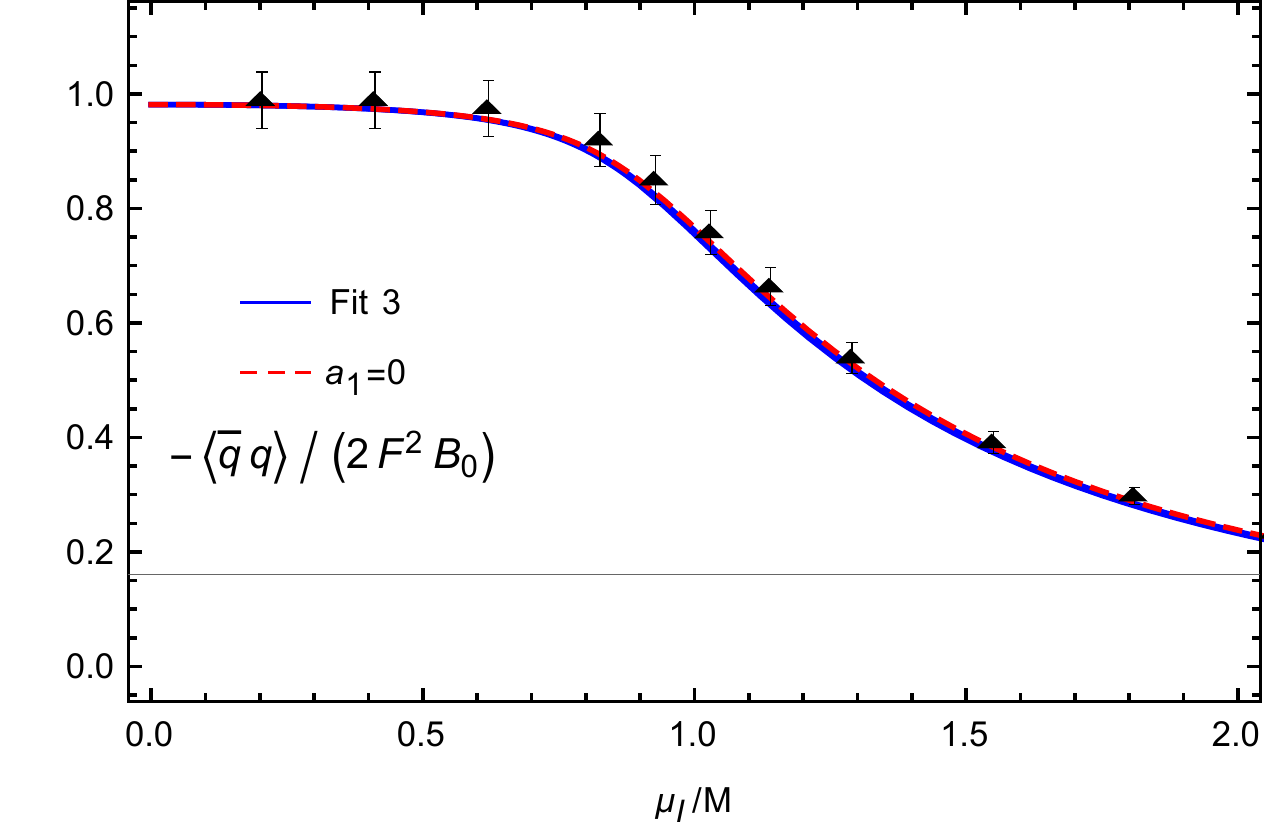}
		\includegraphics[width=0.48\textwidth]{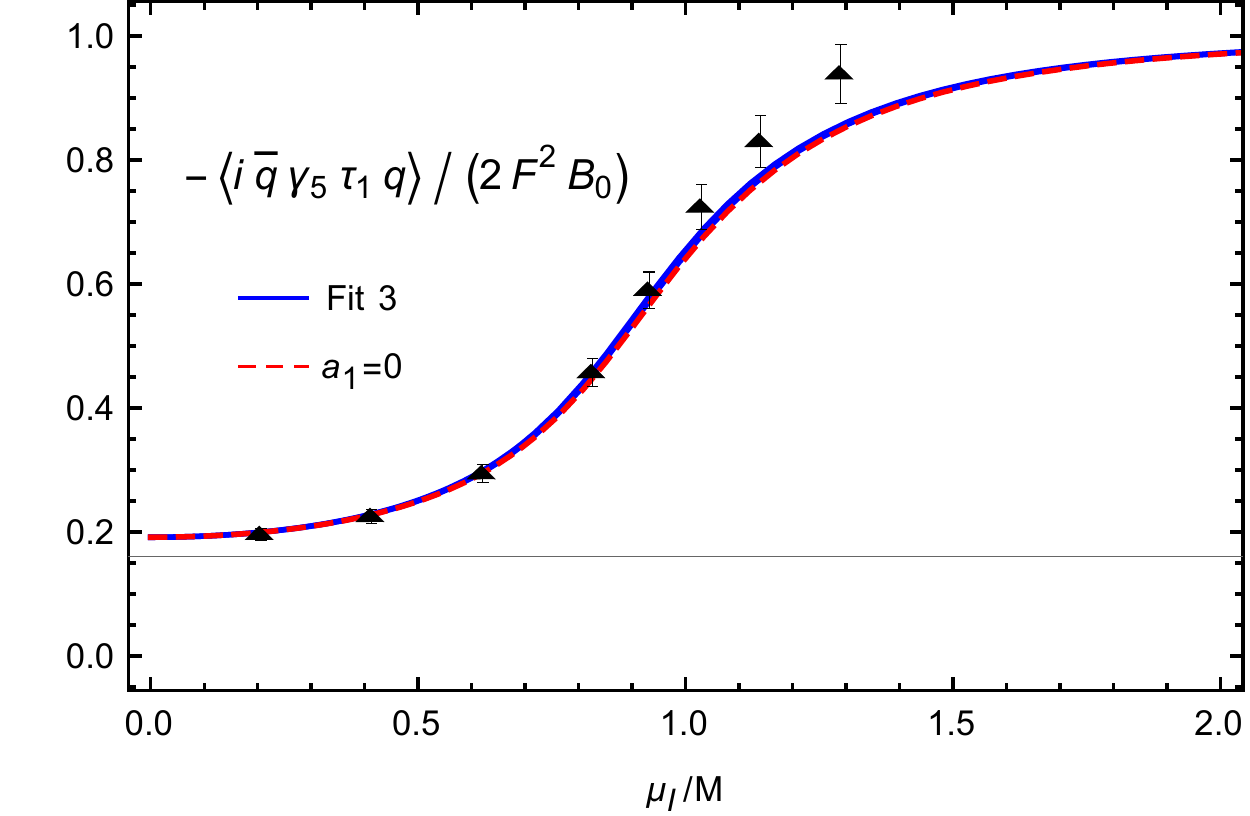}\\
		\includegraphics[width=0.48\textwidth]{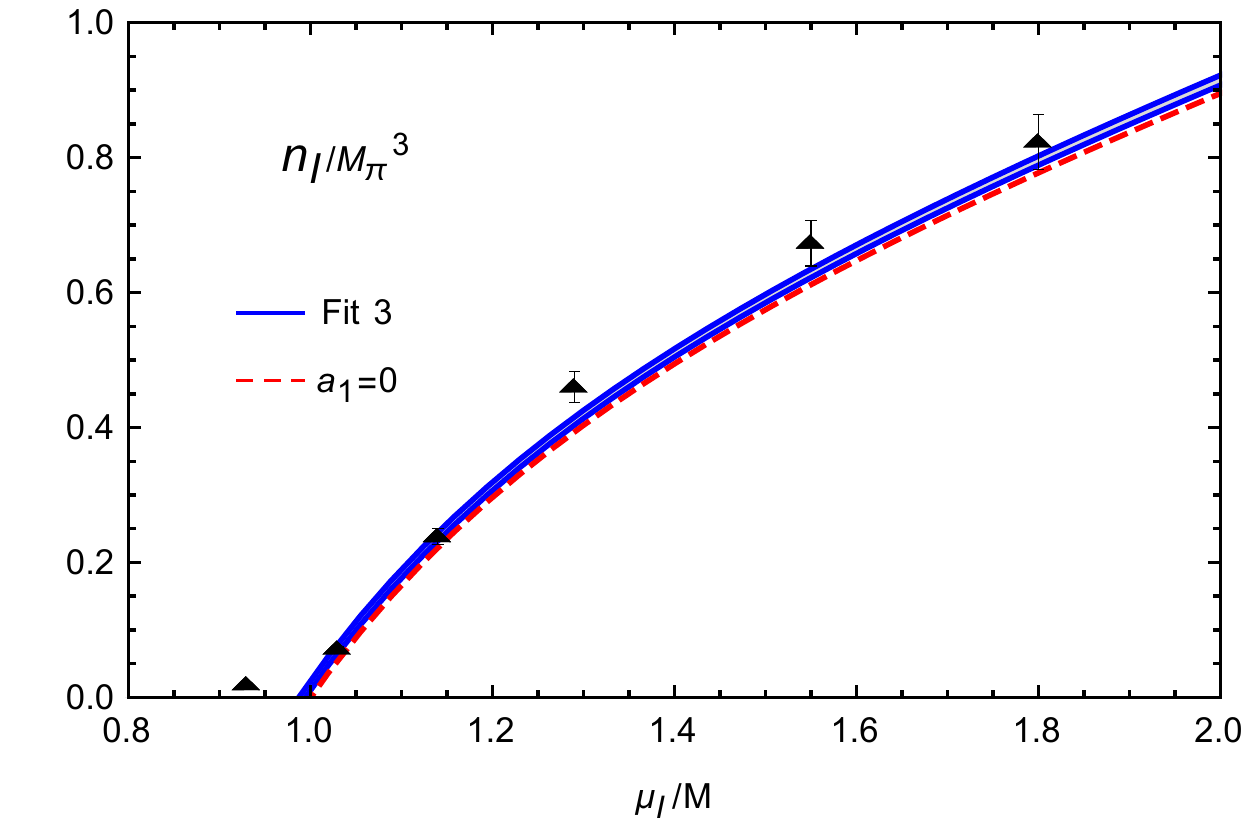}
	\caption{Combined fit of the quark condensate (upper left panel), the pion condensate (upper right panel) and the isospin density (lower panel). Lattice points are the same as in Figs \ref{fig:comfitcondqqpi} and \ref{fig:fitnI}.}
	\label{fig:comfitqqpinI}
\end{figure}

Thus, we show here the results of three different fits, which are summarized in Table \ref{tab:fitall} 
and Figs. \ref{fig:comfitcondqqpi}, \ref{fig:fitnI} and \ref{fig:comfitqqpinI}.  The uncertainty bands in the figures and the uncertainty range for the $a_1$ parameter correspond to the $95\%$ confidence level.

%Due to the uncertainties of the data from \cite{Adhikari:2020ufo} are pretty small, to determinate $a_1$ with reasonable uncertainty we have considered an error of $5\%$ for the condensates while we have used the lattice uncertainty given in \cite{Brandt:2018bwq} for the isospin density. 

Those results lead to the following partial conclusions. First, the quark and pion condensates can be reasonably fitted with $a_1=0$ within the uncertaintites considered (fit 1). Fitting only the quark (pion) condensate favors a positive (negative) $a_1$ value, still compatible with zero. However, the conclusion is very different when fitting the isospin density in fit 2, for which we have  included only the lattice points with $\mu_I>M_\pi$. As explained in  \cite{Brandt:2018bwq}, the lattice uncertainty for $\mu_I\ \sim M_\pi$ is actually much higher, although we have still plotted the first lattice point below $\mu_I$ and given in \cite{Brandt:2018bwq}. As mentioned above, that point lies  below $M_\pi$ with a nonzero isospin density central value, which is actually favored by our present analysis with negative $a_1$.   In fact, as can be seen from the results in Table \ref{tab:fitall}, the lattice results for $n_I$ are much better fitted with a negative value for $a_1$ than for $a_1=0$, as  the values of the corresponding $\chi^2/\text{dof}$ clearly show. The same conclusion is reached when performing a combined fit of the three observables (fit 3), with significant corrections from the $a_1=0$ case. Note also that the fits for $n_I$ reproduce quite well the two ends of the lattice points, i.e., the closest points to $\mu_I=M_\pi$ and those for larger $\mu_I>M_\pi$, improving  over the $a_1=0$ case. 

%On the other hand, when the $\chi^2$ of the quark and pion condensates are combined and minimized at the same time (Table \ref{qqpicomb}), the fit value is zero and the positive and negative uncertainties are similar.

%\begin{table}[h!]
%	\begin{center}
%		\begin{tabular}{|c|c|c||c|}
%			\hline
%			FIT & $a_1$ & $\chi^2/\text{dof}$ & $\chi^2/\text{dof}$, $a_1=0$\\
%			\hline
%			$1: \langle  \bar{q}q\rangle$  & $0.038^{+0.044}_{-0.048}$   &   $0.05$ & 0.28\\
%			\hline
%			$2: \condpiquark$  &  $-0.056^{+0.060}_{-0.061}$  & 0.68  & 1.14 \\
%			\hline
%		\end{tabular}
%		\caption{Numerical values of $a_1$ corresponding to the fit of the quark condensate and the pion condensate respectively. The third column shows the $\chi^2/\text{dof}$ of the fit and the fourth indicates the $\chi^2/\text{dof}$ calculated at $a_1=0$.}
%		\label{qqypi}
%	\end{center}
%\end{table}

\begin{table}[h!]
	\begin{center}
		\begin{tabular}{|l|c|c||c|}
			\hline
			{\bf FIT} & $a_1$ & $\chi^2/\text{dof}$ &  $\chi^2/\text{dof}$, $a_1=0$\\
%			\hline
%			{\bf 1:} $\langle  \bar{q}q\rangle$  & $0.038^{+0.044}_{-0.048}$   &   $0.05$ & 0.28\\
%			\hline
%			{\bf 2:}  $\condpiquark$  &  $-0.056^{+0.060}_{-0.061}$  & 0.68  & 1.14 \\
			\hline
			%{\bf 3:} 
			 {\bf 1:}  $\condpiquark$, $\langle  \bar{q}q\rangle$   &  $0.000^{+0.036}_{-0.038}$  & 0.64  & 0.64 \\
			\hline
			%{\bf 6:}  
			 {\bf 2:}  $n_I$   &  $-0.020\pm 0.007$   &   1.39 & 7.30 \\
			\hline
			%{\bf 4:}  
			{\bf 3:} $\condpiquark$, $\langle  \bar{q}q\rangle$, $n_I$   &  $-0.019\pm 0.007$   &   0.84 & 2.03 \\
			\hline
		\end{tabular}
		\caption{Results for the $a_1$ constant in different fits to quark and pion condensates and isospin density.}
		\label{tab:fitall}
	\end{center}
\end{table}

%\begin{table}[h!]
%	\begin{center}
%		\begin{tabular}{|c|c|c|}
%			\hline
%			FIT 3 & $a_1$ & $\chi^2/\text{dof}$\\
%			\hline
%			$\langle  \bar{q}q\rangle$ and $\condpiquark$ & $0.000^{+0.036}_{-0.038}$   &   $0.64$\\
%			\hline
%		\end{tabular}
%		\caption{Numerical value of $a_1$ corresponding to the combined fit of the quark condensate and the pion condensate.}
%		\label{qqpicomb}
%	\end{center}
%\end{table}

%\begin{figure}[h!]
%	\begin{minipage}{0.48\textwidth}
%		\includegraphics[width=7cm]{qqlattfitlo5.pdf}
%	\end{minipage}
%	\begin{minipage}{0.48\textwidth}
%		\includegraphics[width=7cm]{pilattfitlo5.pdf}
%	\end{minipage}
%	\caption{Fits of the quark condensate (left) and the pion condensate (right). The lattice points used for the fit are those in}
%	\label{fitcondqqpi}
%\end{figure}

%The combined fit of the quark and pion condensates and the isospin density gives a value of $a_1$ which is negative. 
%
%We have fitted the isospin density to lattice data taking into account the points appearing in figure \ref{fitnI} with an isospin chemical potential normalized by $M$ higher than one. We get a similar result for $a_1$ fitting all data. Thus, these fits note that $n_I$ could vanish at $\mu_I/M<1$ and lattice data is in agreement with that.\\ \vspace{0.5cm}

%\begin{table}[h!]
%	\begin{center}
%		\begin{tabular}{|c|c|c||c|}
%			\hline
%			FIT 4 & $a_1$ & $\chi^2/\text{dof}$ & $\chi^2/\text{dof}$, $a_1=0$\\
%			\hline
%			$\langle  \bar{q}q\rangle$, $\condpiquark$ and $n_I$ & $-0.046\pm0.009$   &   $1.60$ & $6.12$\\
%			\hline
%		\end{tabular}
%		\hspace{0.5cm}
%		\begin{tabular}{|c|c|c||c|c|}
%			\hline
%			$n_I$ & $a_1$ & $\chi^2/\text{dof}$& $\chi^2/\text{dof}$, $a_1=0$ & $\#$ points\\
%			\hline
%			FIT 5 & $-0.058^{+0.007}_{-0.009}$   &   $14.50$& $46.53$ & 6\\
%			\hline
%			FIT 6 & $-0.048\pm 0.009$   &   $4.07$ & $26.96$ & 5\\
%			\hline
%		\end{tabular}
%		\caption{Left: Numerical value of $a_1$ corresponding to the combined fit of the quark condensate, the pion condensate and the isospin density. Right: Numerical values of $a_1$ corresponding to the fit of the isospin density with five points (all with $\mu_I/M>1$) and six points (one of them with $\mu_I/M<1$ ).}
%		\label{nIand combcond}
%	\end{center}
%\end{table}

To end this section, we show the results for $\alpha_0^{LO}$, i.e., the solution of the equation $\frac{\partial \epsilon_2}{\partial\alpha}=0$, for the value of $a_1$ obtained in the joint fit 3, both for $j=0$ and for the value of $j$ used here. The explicit expression of $\alpha_0^{LO}$ is given in \eqref{alpha0}. While for $j=0$, the $a_1$ contribution changes the transition point to the BEC phase, as already comented, for $j\neq 0$ the transition is a crossover  and $a_1$ merely modifies the inflection point of the $\alpha_{0}^{LO}$ function.

\begin{figure}[h!]
	\begin{center}
%		\begin{minipage}{0.48\textwidth}
			\includegraphics[width=0.48\textwidth]{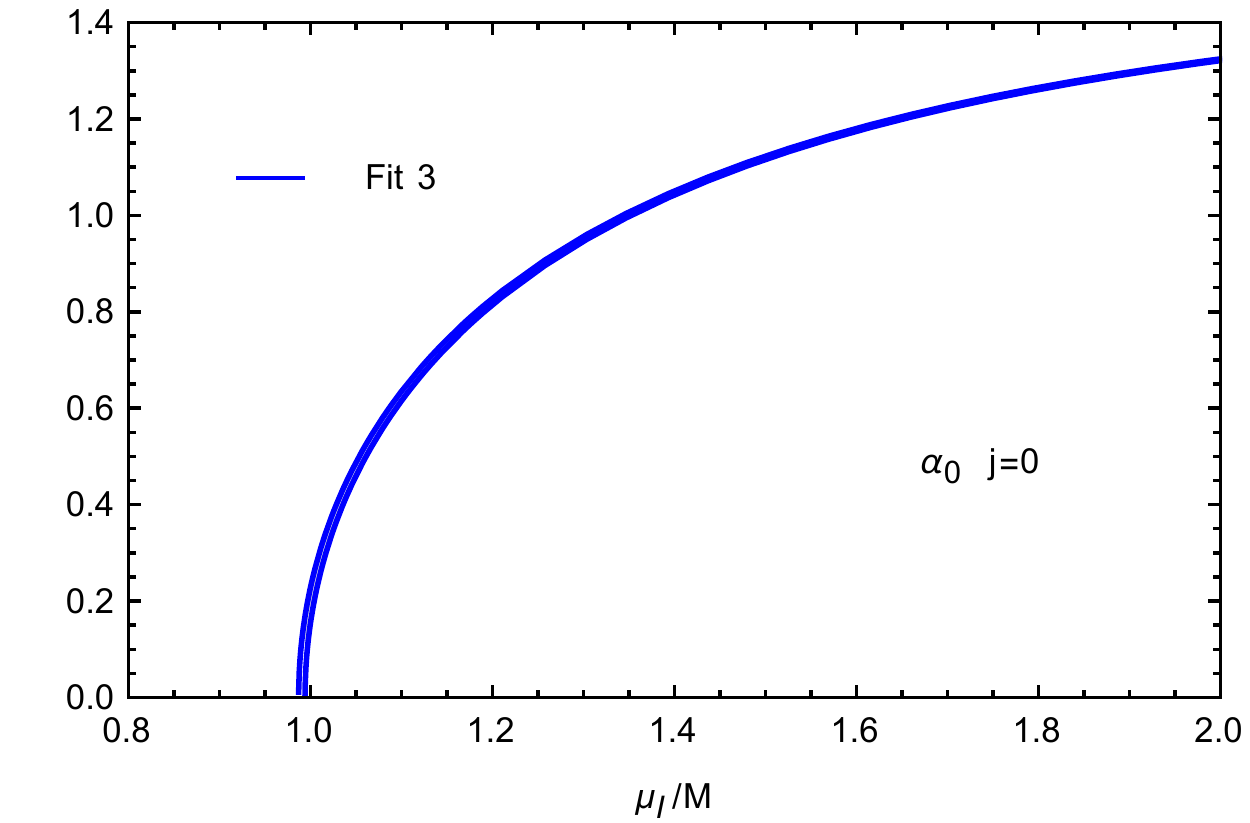}
%		\end{minipage}
%		\begin{minipage}{0.48\textwidth}
			\includegraphics[width=0.48\textwidth]{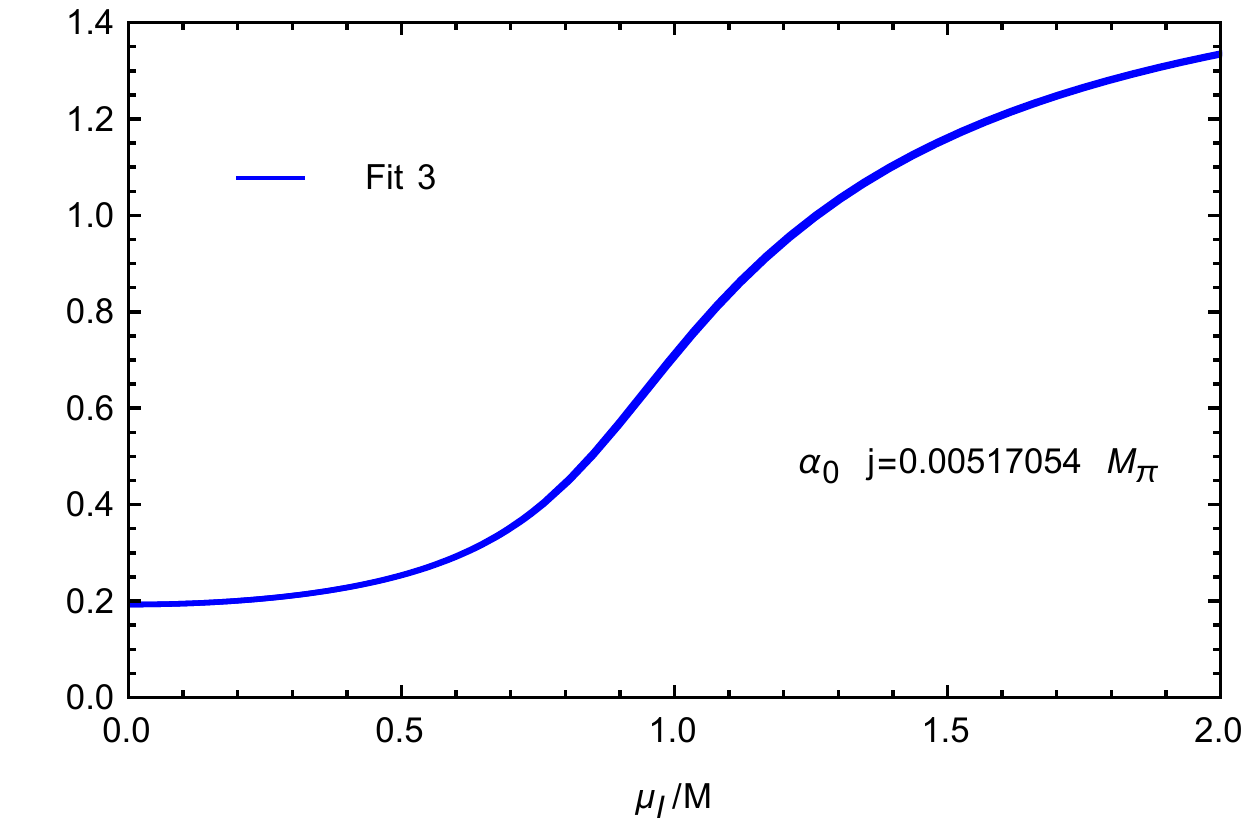}
%		\end{minipage}
	\end{center}
	\caption{$\alpha_{0}^{LO}$ as a function of $\mu_I/M$ for $j=0$ (left) and $j=0.00517054 M_\pi$ (right).}
	\label{alphagsfit}
\end{figure}

To summarize this section, lattice results are perfectly compatible with a coefficient $a_1\neq 0$, which actually improves the description of the isospin density and the overall description of $n_I$ and the two condensates.

\section{Next to leading order results}
\label{sec:nlo}

The next to leading corrections $\epsilon_4$ to the  energy density  come from three different sources:

\begin{itemize}
\item Loop corrections coming from the  quadratic field terms \eqref{L2quad} in the ${\cal L}_2$ lagrangian, which can be obtained in a similar fashion as to the standard free-field contributions to the partition function in vacuum, including now  the $\mu_I$-corrections to the pion dispersion relation \cite{Adhikari:2019mdk}.

\item The constant terms coming from the ${\cal L}_4$ lagrangian in \eqref{L4usual}-\eqref{L4Q}. The LEC coming from this contribution, including the new ones coming from ${\cal L}_4^Q$, will be renormalized to absorb the loop divergences. 

\item The linear field terms coming from ${\cal L}_2$ and given in \eqref{L2lin} will also contribute as long as the  minimizing angle $\alpha_0^{NLO}\neq \alpha_0^{LO}$. This contribution has not been considered in previous  works on this subject and is discussed below.

\end{itemize}

\subsection{Loop contributions} 

Following the same steps as in previous works \cite{Adhikari:2019mdk} we can write the one-loop contribution to the  energy density as

\begin{equation}
\epsilon_4^{loop}=  \frac{1}{2}\int_p \left[E_+ (p) + E_- (p) \right] + \frac{1}{2}\int_p E_0(p)
\label{energy4loop}
\end{equation}
with $E_\pm (p)$ and $E_0(p)$   in \eqref{Epm}-\eqref{E0} and, following the notation in  \cite{Adhikari:2019mdk},
\begin{equation}
\int_p=\mu^{2\epsilon}\int\frac{d^{D-1} p}{(2\pi)^{D-1}}
\end{equation}
with $D=4-2\epsilon$ and $\epsilon\rightarrow 0^+$ and $\mu$ the Dimensional Regularization (DR) scale.

 The treatment of the above integrals, separating their UV divergent contribution in DR follows the same steps as in \cite{Adhikari:2019mdk} except for the modifications proportional to $a_1$ in  equations \eqref{m12}-\eqref{m3sq}. The contribution from the charged pions to the one-loop energy density can be written as
\begin{equation}
\frac{1}{2}\int_p \left[E_+ (p) + E_- (p) \right] =\epsilon_{4,+-}^{div}+\epsilon_{4,+-}^{fin}
\label{energychargedp}
\end{equation}
with
\begin{eqnarray}
\epsilon_{4,+-}^{div}&=&\frac{1}{2}\int_p \left[E_1 (p) + E_2 (p) \right]\label{e4chargdiv}\\
\epsilon_{4,+-}^{fin}&=&\frac{1}{2}\int_p \left[E_+ (p) + E_- (p) - E_1 (p) - E_2 (p) \right]\label{e4chargfin}
\end{eqnarray}
and where we have introduced the quantities $E_{1,2}(p)$ given by
\begin{equation}
\begin{split}
E_{1,2}(p)&=\sqrt{p^2+\hat{m}_{1,2}^2}\\
\hat{m}_1^2&=(M^2+\mu_I^2\cos{\alpha})\cos{\alpha}-(1-a_1)\mu_I^2\cos{2\alpha}+2B_0 j \sin{\alpha}\\
\hat{m}_2^2&=(M^2+a_1\mu_I^2\cos{\alpha})\cos{\alpha}+2B_0 j \sin{\alpha} \\
\hat{m}_3^2&=m_3^2=M^2\cos\alpha +(1-a_1)\mu_I^2\sin^2\alpha+2B_0 j \sin\alpha
\end{split}
\end{equation}

Separating in this way the divergent part of the charged contribution, $\epsilon_{4,+-}^{div}$ in \eqref{e4chargdiv}  has the form of the neutral part in \eqref{energy4loop} and is easier to handle in DR. Actually, note that the large-$p$ behavior of $E_1+E_2$ is the same as the sum $E_+ +E_-$. The finite contribution (subtraction integral) $\epsilon_{4,+-}^{fin}$ can be calculated numerically. Thus, in DR we get for the full divergent part

\begin{equation}
    	\begin{split}
	\left(\epsilon_{4}^{loop}\right)_{div}=&\frac{1}{2}\int_p \left[E_1 (p) + E_2 (p) \right] + \frac{1}{2}\int_p E_0(p)
	% \\
    		%\left(\epsilon_{4}^{loop}\right)_{div}=
		=-\sum_{i=1}^3 \dfrac{\hat{m}_i^4}{4(4\pi)^2}\left[N_\epsilon+\dfrac{3}{2}+\log\left(\dfrac{\mu^2}{\hat{m}_i^2}\right)\right]
%		&-\dfrac{\hat{m}_1^4}{4(4\pi)^2}\left[\dfrac{1}{\epsilon}-\gamma+\log(4\pi)+\dfrac{3}{2}+\log\left(\dfrac{\mu^2}{\hat{m}_1^2}\right)\right]\\
%    		&-\dfrac{\hat{m}_2^4}{4(4\pi)^2}\left[\dfrac{1}{\epsilon}-\gamma+\log(4\pi)+\dfrac{3}{2}+\log\left(\dfrac{\mu^2}{\hat{m}_2^2}\right)\right]\\
%    		&-\dfrac{m_3^4}{4(4\pi)^2}\left[\dfrac{1}{\epsilon}-\gamma+\log(4\pi)+\dfrac{3}{2}+\log\left(\dfrac{\mu^2}{m_3^2}\right)\right]
    	\end{split}
    \label{energydivloop}
\end{equation}
with $N_\epsilon=\dfrac{1}{\epsilon}-\gamma+\log(4\pi)$ and $\gamma$ the Euler constant. 

\subsection{Fourth order lagrangian and renormalization} 

The contributions to the energy density coming from the constant part in the fourth order lagrangian \eqref{L4usual} and \eqref{L4Q}  are respectively given by

\begin{eqnarray}
\epsilon_4^{40}&=&-\left(l_1-l_2\right)\mu_I^4\sin^4\alpha-l_4M^2\mu_I^2 \cos\alpha+l_4M^2\mu_I^2 \cos^3\alpha-2l_4B_0\mu_I^2 j\sin\alpha+2l_4B_0\mu_I^2 j\sin\alpha\cos^2\alpha\nonumber\\
&-&\left(h_1-l_4\right)M^4-(l_3+l_4)M^4\cos^2\alpha-2B_0 j (l_3+l_4)M^2\sin( 2\alpha)-4B_0^2j^2\left[h_1-l_4+(l_3+l_4)\sin^2\alpha\right]
\label{energy40}\\
\epsilon_4^{4Q}&=&-\hat{q}_1\mu_I^4\sin^4\alpha-\frac{1}{2}\hat{q}_2M^2\mu_I^2 \cos\alpha-\frac{1}{2}\hat{q}_3M^2\mu_I^2 \cos^3\alpha-\hat{q}_4B_0\mu_I^2 j\sin\alpha-\hat{q}_5B_0\mu_I^2 j\sin\alpha\cos^2\alpha\nonumber\\
&-&\hat{q}_6\mu_I^4\cos^2\alpha-\hat{q}_7\mu_I^4
\label{energy4Q}
\end{eqnarray}
with
\begin{align*}
	\hat{q}_1&=q_{10}-2q_2-2q_3-q_4&\hat{q}_5&=8(q_6+q_7)\\
	\hat{q}_2&=4q_5-2q_6-8q_7&\hat{q}_6&=\dfrac{1}{2}(-2q_1+2q_{10}-2q_2+q_9)\\
	\hat{q}_3&=4(q_6+2q_7)&\hat{q}_7&=\dfrac{1}{4}(4q_1-3q_{10}+4q_2+q_8-q_9)\\
	\hat{q}_4&=4(q_5-q_6)
\end{align*}

Comparing \eqref{energy4Q} with \eqref{energy40}, we see that new terms introduce $\mu_I$-dependent corrections as follows: $\hat{q}_1$ shifts the $(l_1-l_2)\mu_I^4$ contribution, $\hat{q}_{2,3,4,5}$ modify respectively the four  $l_4\mu_I^2$ terms, whereas $\hat{q}_{6,7}$  introduce new $\mu_I^6$ terms. Of those new seven independent LECs appearing in the energy density, $\hat{q}_4$ and $\hat{q}_5$ will not contribute for $j=0$ and the $\hat{q}_7$ term is independent of $\alpha$ so it does not contribute to the energy density minimum.

The new LECs will precisely absorb the new loop divergences dependent on $a_1$ included in \eqref{energydivloop}. Thus, the renormalized energy density at NLO resulting of adding all the contributions mentioned before can be written as
    
	\begin{equation}
	\small{
		\begin{split}
    		&\epsilon_4^{loop}+\epsilon_4^{40}+\epsilon_4^{4Q}=-\dfrac{1}{(4\pi)^2}\left[\dfrac{3}{2}-\bar{l}_3+4\bar{l}_4+\log\left(\dfrac{M^2}{\hat{m}_2^2}\right)+2\log\left(\dfrac{M^2}{\hat{m}_3^2}\right)\right]\left[\frac{1}{4}M^4\cos^2\alpha+B_0^2j^2\sin^2\alpha+\frac{1}{2}M^2B_0j\sin(2\alpha)\right]\\
		&-\dfrac{1}{2(4\pi)^2}\left\lbrace\dfrac{1}{2}+\dfrac{1}{3}\bar{l}_1+\dfrac{2}{3}\bar{l}_2+2(4\pi)^2\hat{q}_1^r-\dfrac{a_1}{2}\left[3(1-a_1)+4(1-a_1)\log\left(\dfrac{\mu^2}{\hat{m}_1^2}\right)-a_1\log\left(\dfrac{\mu^2}{\hat{m}_2^2}\right)+(2-a_1)\log\left(\dfrac{\mu^2}{m_3^2}\right)\right]\right.\\
    		&\left.
		% \left.+6(1-a_1)\log\left(\dfrac{\mu^2}{M^2}\right)\right]
		+\log\left(\dfrac{M^2}{m_3^2}\right)\right\rbrace\mu_I^4\sin^4\alpha \\
    		&-\dfrac{1}{(4\pi)^2}\left\lbrace\dfrac{1}{2}+\bar{l}_4+\dfrac{(4\pi)^2}{2}\hat{q}_2^r-\frac{a_1}{2}\left[1+\log\left(\dfrac{\mu^2}{\hat{m}_1^2}\right)+\log\left(\dfrac{\mu^2}{m_3^2}\right)
		%+4\log\left(\dfrac{\mu^2}{M^2}\right)
		\right]+\log\left(\dfrac{M^2}{m_3^2}\right)\right\rbrace M^2\mu_I^2\cos\alpha\\
    		&+\dfrac{1}{(4\pi)^2}\left\lbrace\dfrac{1}{2}+\bar{l}_4-\dfrac{(4\pi)^2}{2}\hat{q}_3^r-\frac{a_1}{2}\left[2+2\log\left(\dfrac{\mu^2}{\hat{m}_1^2}\right)+\log\left(\dfrac{\mu^2}{\hat{m}_2^2}\right)+\log\left(\dfrac{\mu^2}{m_3^2}\right)
		%+4\log\left(\dfrac{\mu^2}{M^2}\right)
		\right]+\log\left(\dfrac{M^2}{m_3^2}\right)\right\rbrace M^2\mu_I^2\cos^3\alpha\\
    		&-\dfrac{2}{(4\pi)^2}\left\lbrace\dfrac{1}{2}+\bar{l}_4+\dfrac{(4\pi)^2}{2}\hat{q}_4^r-\frac{a_1}{2}\left[1+\log\left(\dfrac{\mu^2}{\hat{m}_1^2}\right)+\log\left(\dfrac{\mu^2}{m_3^2}\right)
		%+4\log\left(\dfrac{\mu^2}{M^2}\right)
		\right]+\log\left(\dfrac{M^2}{m_3^2}\right)\right\rbrace B_0j\mu_I^2\sin\alpha\\
    		&+\dfrac{2}{(4\pi)^2}\left\lbrace\dfrac{1}{2}+\bar{l}_4-\dfrac{(4\pi)^2}{2}\hat{q}_5^r-\frac{a_1}{2}\left[2+2\log\left(\dfrac{\mu^2}{\hat{m}_1^2}\right)+\log\left(\dfrac{\mu^2}{\hat{m}_2^2}\right)+\log\left(\dfrac{\mu^2}{m_3^2}\right)
		%+4\log\left(\dfrac{\mu^2}{M^2}\right)
		\right]+\log\left(\dfrac{M^2}{m_3^2}\right)\right\rbrace B_0j\mu_I^2\sin\alpha\cos^2\alpha\\
    	%	&-\dfrac{1}{2(4\pi)^2}\left\lbrace\dfrac{1}{2}+\dfrac{1}{3}\bar{l}_1+\dfrac{2}{3}\bar{l}_2+2(4\pi)^2\hat{q}_1^r-\dfrac{a_1}{2}\left[3(1-a_1)+4(1-a_1)\log\left(\dfrac{M^2}{\hat{m}_1^2}\right)-a_1\log\left(\dfrac{M^2}{\hat{m}_2^2}\right)+(2-a_1)\log\left(\dfrac{M^2}{m_3^2}\right)\right.\right.\\
    		%&
		%\left.\left.+6(1-a_1)\log\left(\dfrac{\mu^2}{M^2}\right)\right]+\log\left(\dfrac{M^2}{m_3^2}\right)\right\rbrace\mu_I^4\sin^4\alpha
		%-\dfrac{1}{2(4\pi)^2}\left\lbrace 2(4\pi)^2\hat{q}_7^r +\dfrac{a_1}{2}\left[1-2a_1+(2-3a_1)\log\left(\dfrac{M^2}{\hat{m}_1^2}\right)-a_1\log\left(\dfrac{M^2}{\hat{m}_2^2}\right)\right.\right.\\
    		%&\left.\left.+2(1-2a_1)\log\left(\dfrac{\mu^2}{M^2}\right)\right]\right\rbrace \mu_I^4
		&-\dfrac{1}{2(4\pi)^2}\left\lbrace2(4\pi)^2\hat{q}_6^r-\dfrac{a_1}{2}\left[1-3a_1+2(1-2a_1)\log\left(\dfrac{\mu^2}{\hat{m}_1^2}\right)-2a_1\log\left(\dfrac{\mu^2}{\hat{m}_2^2}\right)
		%\right.\right.
		%\\
    		%&\left.\left.
		%+2(1-3a_1)\log\left(\dfrac{\mu^2}{M^2}\right)
		\right]\right\rbrace \mu_I^4\cos^2\alpha \\
		&-\dfrac{1}{2(4\pi)^2}\left\lbrace 2(4\pi)^2\hat{q}_7^r +\dfrac{a_1}{2}\left[1-2a_1+(2-3a_1)\log\left(\dfrac{\mu^2}{\hat{m}_1^2}\right)-a_1\log\left(\dfrac{\mu^2}{\hat{m}_2^2}\right)
		%\right.\right.\\
    		%&\left.\left.
		%+2(1-2a_1)\log\left(\dfrac{\mu^2}{M^2}\right)
		\right]\right\rbrace \mu_I^4 
		%\\
		%&
		-\dfrac{1}{(4\pi)^2}(\bar{h}_1-\bar{l}_4)\left[M^2+4B_0^2j^2\right]+\epsilon_{4,+-}^{fin}
    	\end{split}}
		\label{energy4qsa1ren}
	\end{equation}
	where the renormalized and scale-independent $\bar l_i,\bar h_i$ are the standard ones given in  \cite{Gasser:1983yg,Adhikari:2020ufo} while the new LECs are renormalized as
	
	\begin{equation}
\hat{q}_i=\hat{q}^r_i(\mu)-\eta_i\frac{\mu^{-2\epsilon}}{2\left(4\pi\right)^2}\left[N_\epsilon+1\right]
\label{qren}
\end{equation}
with 

\begin{align}
	\eta_1&=-3 a_1+3a_1^2&\eta_5&=8a_1 \nonumber\\
	\eta_2&=-4a_1&\eta_6&=-a_1+3a_1^2 \nonumber\\
	\eta_3&=8a_1&\eta_7&=a_1-2a_1^2 \nonumber\\
	\eta_4&=-4a_1
	\label{qrencoeff}
\end{align} 
where, as usual, the $\mu$-dependence of the renormalized LECs cancels with that of the loops,  rendering the energy density finite and scale independent. In the following sections we will analyze the effect of these new LECs on the $\mu_I$-dependence of the different observables obtained from the energy density. 

The numerical values we will use for the $\bar l_i$ as well as for the physical pion mass and decay constant will be the same as the  previous works on this subject \cite{Adhikari:2019zaj,Adhikari:2020ufo} for an easier comparison. Thus, we will take the central values of the $\bar l_i,\bar h_i$ from \cite{Colangelo2001}:

\begin{equation}
    \bar{l}_1=-0.4,\hspace{0.5cm}\bar{l}_2=4.3,\hspace{0.5cm}\bar{l}_3=2.9,\hspace{0.5cm}\bar{l}_4=4.4,\hspace{0.5cm}\bar{h}_1-\bar{l}_4=-1.5,
\label{lecslh}
\end{equation}
which have been used to study the quark and pion condensates in \cite{Adhikari:2020ufo} and the isospin density in \cite{Adhikari:2019zaj} at zero temperature. 

As for the physical $M_\pi,F_\pi$, as commented above, we are considering those used in the lattice simulations, which using the one-loop standard ChPT expressions \cite{Gasser:1983yg} give rise for the following values of the tree-level $M,F$ appearing in our previous expressions for the energy density:

\begin{equation}
M_\pi=131 \ \text{MeV},\hspace{0.5cm}M=132.49 \ \text{MeV},\hspace{0.5cm} F_\pi=90.51 \ \text{MeV},\hspace{0.5cm}F=84.93 \ \text{MeV}.
\label{massfpi}
\end{equation}

Recall that the $\bar l_i, \bar h_i$, although scale-independent are mass-dependent and therefore there might be slight numerical variations from the values \eqref{lecslh} when taking  the masses in \eqref{massfpi}. Those variations are logarithmic and therefore numerically negligible, so for our purposes  of comparing with previous works we will still use the values in \eqref{lecslh}.

\subsection{Linear terms} 
\label{sec:lin}

The solution  for the angle minimizing the energy density to leading order, given in  \eqref{alpha0} for $j=0$, is such that the linear term in \eqref{L2lin} proportional to $\pi_1(x)$ vanishes so the linear terms can be ignored since the derivative term in \eqref{L2lin} does not contribute to the vacuum energy density. However, this is not neccessarily true to higher orders. That is, if we consider a perturbative deviation of the LO minimizing angle,

\begin{equation}
\alpha=\alpha_0^{LO}+\delta\alpha_0
\label{angledev}
\end{equation}
with $\alpha_0^{LO}=\Od(1)$, $\delta\alpha_0=\Od(1/F)$ in the chiral expansion, we can write

\begin{equation}
{\cal L}_2^{lin}=f'(\alpha_0^{LO}) (\alpha-\alpha_0^{LO}) \pi_1 (x) + \Od(1/F) + \dots
\label{L2linexpanded}
\end{equation}
where the dots denote derivative terms and, according to \eqref{L2lin}, 

\begin{equation}
f(\alpha)=-F\sin\alpha\left[M^2- (1-a_1)\mu_I^2\cos\alpha\right]+2B_0 F j  \cos\alpha
\label{ffun}
\end{equation}
 
 The above linear contribution to NLO can be reabsorbed into a  redefinition of the $\pi_1$ field, over which we are integrating to get the energy density. Namely,
 
 \begin{equation}
 \pi_1\rightarrow \pi_1+\frac{f'\left(\alpha_0^{LO}\right)}{m_1^2\left(\alpha_0^{LO}\right)}(\alpha-\alpha_0^{LO})
 \end{equation}

The above shift eliminates the linear term at this order and completing squares generates the following additional $\Od(1)$ perturbative contribution to the NLO energy density:

\begin{equation}
\epsilon_4^{lin}=-\frac{1}{2} \frac{\left[f'\left(\alpha_0^{LO}\right)\right]^2}{m_1^2\left(\alpha_0^{LO}\right)}\left(\alpha-\alpha_0^{LO}\right)^2
\label{energy4linear}
\end{equation}
with $m_1^2$ in \eqref{m1sq}, which added to the  contributions given in \eqref{energy4loop}, \eqref{energy40} and \eqref{energy4Q} 
has to be minimized with respect to $\alpha$, around $\alpha_0^{LO}$, to find $\alpha_0^{NLO}$. Moreover, it must be taken into account to calculate different observables because of its $m$, $j$ and $\mu_I$ dependences:
\begin{eqnarray}
\Delta\langle \bar{q}q\rangle^{NLO}_{lin}&=&\dfrac{\partial\epsilon_4^{lin}}{\partial m}=-B_0 F^2\cos(\alpha_0^{LO})\left(\alpha-\alpha_0^{LO}\right)^2\label{qqlin}\\
\Delta\langle i\bar{q}\gamma_5\tau_1q\rangle^{NLO}_{lin}&=&\dfrac{\partial\epsilon_4^{lin}}{\partial j}=-B_0 F^2\sin(\alpha_0^{LO})\left(\alpha-\alpha_0^{LO}\right)^2\label{pilin}\\
\Delta (n_I)^{NLO}_{lin}&=&-\dfrac{\partial\epsilon_4^{lin}}{\partial \mu_I}=-\mu_I F^2\left(1-a_1\right)\cos(2\alpha_0^{LO})\left(\alpha-\alpha_0^{LO}\right)^2\label{nIlin}
\end{eqnarray}

\begin{figure}[h!]
  \centering{
  \includegraphics[width=0.48\textwidth]{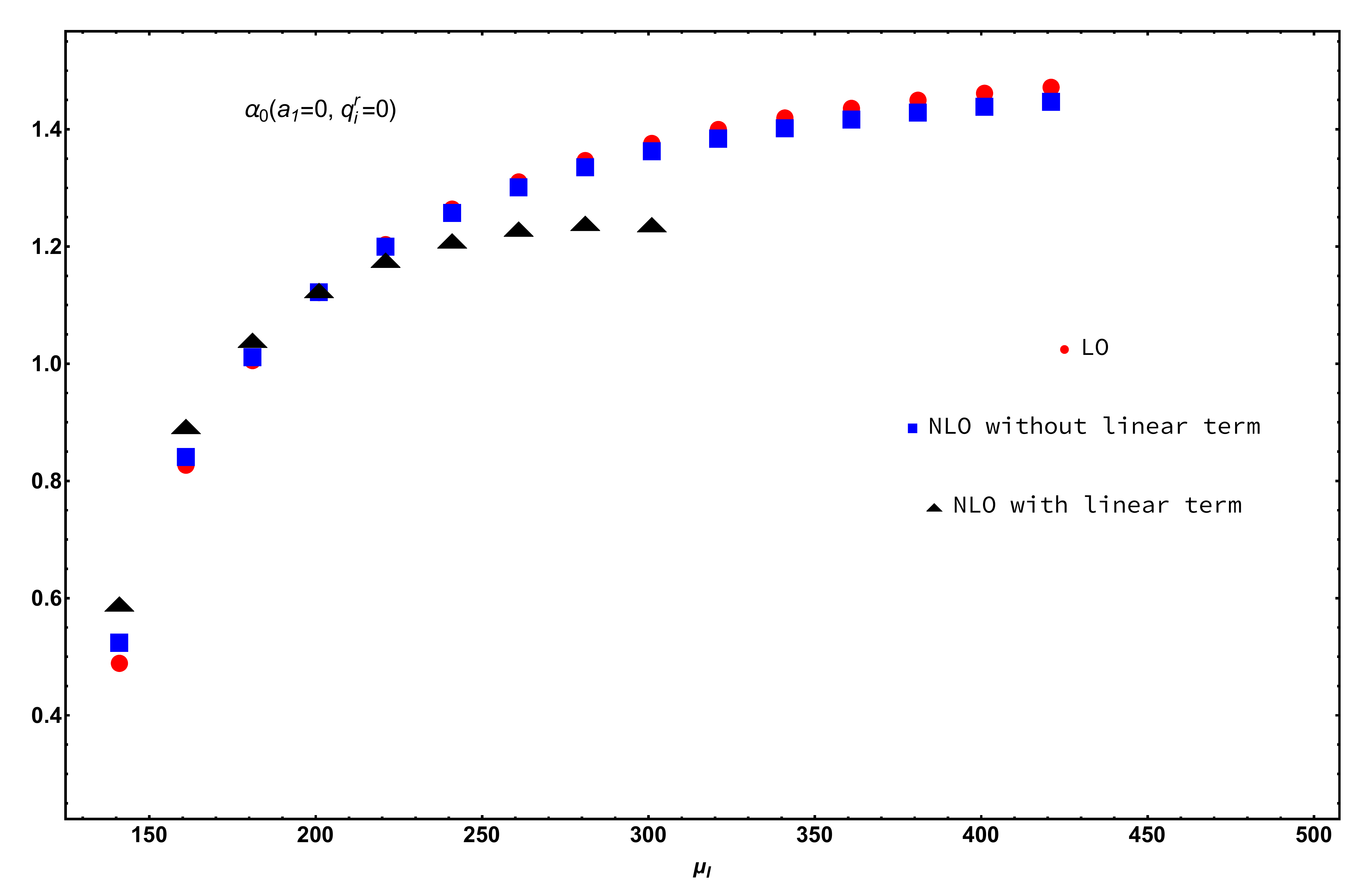}
  \includegraphics[width=0.47\textwidth]{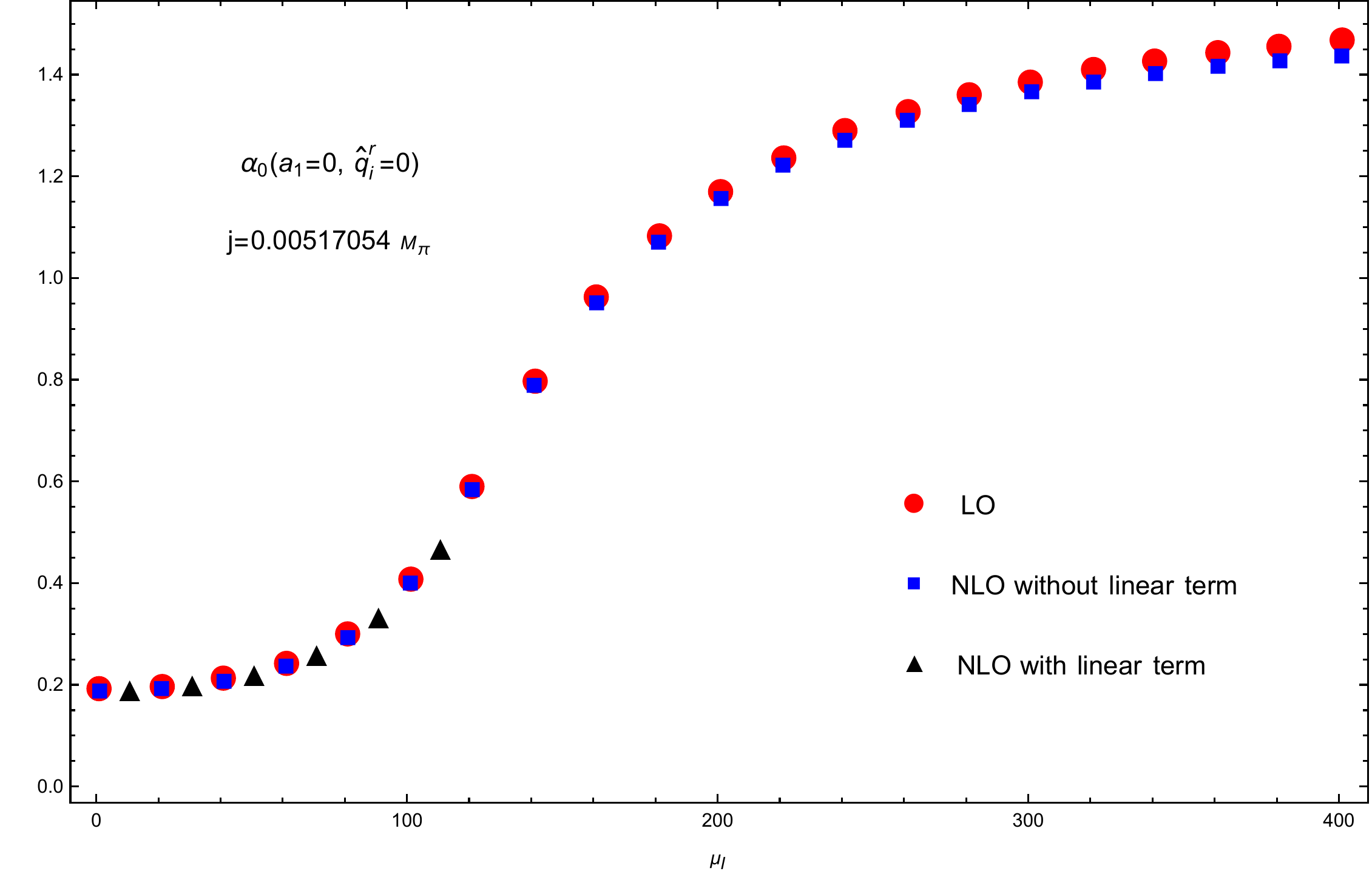}}
  \caption{Effect of the linear term in the minimizing angle $\alpha_0$  for $j=0$ (left) and $j\neq 0$ (right).}
  \label{Figure:angle}
\end{figure}

To have a more quantitative idea of the effect of this correction, we have plotted in Fig. \ref{Figure:angle} the result for the minimizing angle $\alpha_0$, comparing the LO contribution with the NLO with and without including the linear term contribution \eqref{energy4linear}.  For easier comparison with previous works, we have not included in that plot  the new contributions coming from the $a_1$ and $q_i$ terms and we have used the same LEC as in \cite{Adhikari:2019zaj}.  We consider both the $j=0$ and $j\neq 0$ situations.   As we can see in that figure, the inclusion of the linear term may generate sizable differences between the NLO and LO results. Actually, for some values of the constants involved, those linear corrections can be such that the effective potential  stops having a minimum above a certain $\mu_I$ value. We can actually see this behaviour  in the plot showed in Fig. \ref{Figure:angle}  for  which that limiting value is $\mu_I\simeq$ 300 MeV  for $j=0$ and $\mu_I\simeq$ 340 MeV  for the $j\neq 0$ value considered.  The deviations with respect to the LO indicate that we are reaching the borderline of the  ChPT validity limit where in particular the very same approximation followed in \eqref{angledev} and \eqref{L2linexpanded} would fail. This is actually consistent   with  lattice analyses showing  that deviations from ChPT around those $\mu_I$ values signal the onset of the BCS phase \cite{Detmold:2012wc}.  Nevertheless, it should be taken into account that the actual value of such validity limit  depends also on the rest of the LEC involved, as we will discuss in detail below.

\subsection{Numerical results to NLO}
\label{sec:numres}

\subsubsection{Constraints on the LEC from the isospin density}
\label{sec:const}

First, let us discuss the constraints that arise for the LEC  from the condition of vanishing isospin density for $\mu_I<\mu_c$, i.e., the extension to NLO of the LO constraint \eqref{locons}. Thus, let us denote 
$\mu_c^{LO} =\frac{M}{\sqrt{1-a_1}}$ and $\mu_c^{NLO}$ the NLO value for $\mu_c$ which will be determined explicitly in section \ref{sec:mucnlo} below and which depends on the $\hat q_i^r$ constants. 

Note that by definition, $\alpha^{NLO}(\mu_I<\mu_c^{NLO})=0$.  In addition, the linear term correction \eqref{energy4linear} vanishes for $\mu_I$ below $\mu_c^{NLO}$ {\em and} $\mu_c^{LO}$  since $\alpha^{LO}(\mu_I<\mu_c^{LO})=0$ but it would survive for $\mu_c^{LO}<\mu_I<\mu_c^{NLO}$ provided $\mu_c^{LO}<\mu_c^{NLO}$. Therefore, taking the derivative with respect to $\mu_I$ and setting $j=\alpha=0$ in our previous expressions for the energy density, we get, 

\begin{itemize}
\item  $\mu_c^{LO}>\mu_c^{NLO}\Rightarrow n_I(\mu_I<\mu_c^{NLO})=n_0 (\mu_I)$
\item $\mu_c^{LO}<\mu_c^{NLO}\Rightarrow \displaystyle n_I(\mu_I<\mu_c^{NLO})=\left\{
\begin{array}{cl}
n_0(\mu_I)+n_1 (\mu_I) & \mu_c^{LO}<\mu_I<\mu_c^{NLO} \\ \\
n_0(\mu_I)&  \mu_I<\mu_c^{LO}\\
\end{array}
\right.
$
\end{itemize}

with

\begin{equation}
    	\begin{split}
    		n_0(\mu_I)&=\dfrac{F^2}{2}(a_1+a_2)\mu_I+M^2\left(\hat{q}_2^r+\hat{q}_3^r\right)\mu_I+4\left(\hat{q}_6^r+\hat{q}_7^r\right)\mu_I^3\\
		&+\dfrac{a_1}{8\pi^2(M^2+a_1\mu_I^2)}\left[\dfrac{M^4}{4}\mu_I+(M^2+a_1\mu_I^2)^2\log\left(\frac{\mu^2}{M^2+a_1\mu_I^2}\right)\right]+\mathcal{O}\left(\frac{1}{F^2}\right)
		%n_I(\mu_I)|_{\alpha=0}=M^2\left[\hat{q}_2^r+\hat{q}_3^r+\dfrac{a_1}{32\pi^2}+\dfrac{a_1}{8\pi^2}\log\left(\dfrac{\mu^2}{M^2}\right)\right]\mu_I+4\left(\hat{q}_6^r+\hat{q}_7^r\right)\mu_I^3+\mathcal{O}(a_1^2)
    	%	&-(1-a_1)(\alpha_0^{LO})^2\cos(2\alpha_0^{LO})\mu_I+\mathcal{O}(a_1^3)
	\\
	n_1(\mu_I)&=\dfrac{1}{2}\dfrac{\partial}{\partial \mu_I}   \left[\dfrac{\alpha_0^{LO}(\mu_I) f_0'\left[\alpha_0^{LO}(\mu_I) \right]}{m_1 (\mu_I)}\right]^2 +\mathcal{O}\left(\frac{1}{F^2}\right)
    	\end{split}
	    \label{nIjalpha0q}
\end{equation}
where $f_0=f(j=0)$ with $f$  in \eqref{ffun} and we have followed \cite{Adhikari:2019mdk} for the calculation of the loop integrals. 

Thus, since the $\mu_I$ dependence of the $n_1$ function above (coming from the linear term) is non-polynomical, the only way to ensure that $n_I$ vanishes for all $\mu_I$ below $\mu_c^{NLO}$ is that the $\hat q_i^r$ satisfy the constraint

\begin{equation}
\mu_c^{NLO} (\hat q_i^r)<\mu_c^{LO}
\label{extracond}
\end{equation}

On the other hand, the condition that the $n_0$ function in \eqref{nIjalpha0q} vanishes,  together with  \eqref{locons}, implies at this order:

\begin{eqnarray}
    	a_1,a_2&=&\Od\left(\frac{1}{F^2}\right)
	\label{ordera1a2}\\
	a_1+a_2&=& -\frac{2M^2}{F^2}\left[\hat{q}_2^r(\mu)+\hat{q}_3^r(\mu)\right]+\Od\left(\frac{1}{F^4}\right)
    	%\hat{q}_3^r(\mu)&=&-\hat{q}_2^r(\mu)-\dfrac{1}{2}\dfrac{F^2}{M^2}a_1
    	\label{q3q2}\\
    \hat{q}_6^r\left(\mu\right) +	\hat{q}_7^r\left(\mu\right)&=&0+\Od\left(\frac{1}{F^2}\right)
	\label{condq7}
    \end{eqnarray}

%\begin{eqnarray}
%a_2&=&-a_1-\dfrac{M^2}{F^2}\left\lbrace2\left[\hat{q}_2^r\left(\mu\right)+\hat{q}_3^r\left(\mu\right)\right]+\dfrac{a_1}{16\pi^2}+\dfrac{a_1}{4\pi^2}\log\left(\dfrac{\mu^2}{M^2}\right)\right\rbrace+\mathcal{O}(a_1^3)\label{q3q2}\label{conda2}\\
%\hat{q}_7^r\left(\mu\right)&=&-\hat{q}_6^r\left(\mu\right)+\dfrac{a_1^2}{32\pi^2}\left[\dfrac{5}{4}-\log\left(\dfrac{\mu^2}{M^2}\right)\right]+\mathcal{O}(a_1^3)
%\label{condq7}
%\end{eqnarray}
%

The  constraints \eqref{q3q2} and \eqref{condq7} allow us to eliminate the dependence on two of the LEC involved in terms of the remaining ones. In fact, note that equation \eqref{q3q2} is actually the explicit expression for the $\Od(1/F^2)$ in \eqref{locons}. We also remark that the condition \eqref{ordera1a2} is fully consistent with having obtained a numerical value $a_1\ll 1$ in our fit study in section \ref{sec:LOfits}. The situation is similar to the EM corrections in ChPT, where there are operators which are formally $\Od(p^2)$, such as  \eqref{examplenew}, but multiplied by $e^2$ which is a numerically small parameter. 

On the other hand, we must keep in mind the constraint \eqref{extracond} when considering the possible numerical range for the $\hat q_i^r$ in the following sections.

\subsubsection{The critical BEC value at NLO}
\label{sec:mucnlo}

Now, let us evaluate the effect of the NLO corrections to the critical value $\mu_{c}$. Since $\frac{\partial\epsilon^{NLO}}{\partial\alpha}|_{\alpha=0}=0$, $\mu_c$ can be determined as the value for which $\alpha=0$ flips from a local minimum to a local maximum. Therefore, expanding the energy density around $\alpha=0$,

\begin{equation}
    \epsilon=\beta_0(\mu_I)+\beta_2(\mu_I)\alpha^2+\mathcal{O}\left(\alpha^4\right)
\end{equation}

we have

\begin{equation}
    \left. \beta_2(\mu_I)\right|_{\mu_I=\mu_c}=0
\end{equation}

From the energy density discussed above, we find

\begin{equation}
	%\begin{split}
		\beta_2(\mu_I)=\dfrac{1}{2}F_\pi^2\left[M_\pi^2-(1-a_1)\mu_I^2\right]+\dfrac{1}{4}\mu_I^2\left[M^2\left(\hat{q}_2^r+3\hat{q}_3^r\right)+4\hat{q}_6^r\mu_I^2\right] -\dfrac{F^2 \left(\mu_I^4-M^4\right)}{2\mu_I^2}
		+\mathcal{O}\left(\frac{1}{F^2}\right)
		%+\dfrac{1}{(M^2+a_1\mu_I^2)}\left\lbrace\dfrac{a_1}{128\pi^2}\left[2\mu_I^2+(1-8\bar{l}_4)M^2\right]M^2\mu_I^2\right.\\
		%&\left. -\dfrac{a_1^2}{64\pi^2}\left(1+4\bar{l}_4\right)\mu_I^4M^2\right\rbrace+\dfrac{1}{64\pi^2}\left[M^4\log\left(\frac{M^2}{M^2+a_1\mu_I^2}\right)+2a_1\mu_I^2M^2\log\left(\dfrac{\mu^2}{M^2}\right)\right.\\
		%&\left.+2a_1\mu_I^2\left[4M^2+(3a_1-1)\mu_I^2\right]\log\left(\frac{\mu^2}{M^2+a_1\mu_I^2}\right)\right]
		%\\
		%&-\dfrac{F^2}{2}\left[M^2\cos(\alpha_0^{LO})-(1-a_1)\mu_I^2\cos(2\alpha_0^{LO})\right]+\mathcal{O}\left(\frac{1}{F^2}\right)
	%\end{split}
\label{beta2}
\end{equation}
where $M_\pi$ and $F_\pi$ are the NLO pion decay constant and mass respectively \cite{Gasser:1983yg} and where we have made use of the condition \eqref{ordera1a2} so that the $a_1$ dependence in the NLO has been ignored.

%As we have seen in previous sections, the lattice results are compatible with a small nonzero value for $a_1$. Thus, for simplicity, we provide here the expression for $\beta_2$ expanding around $a_1=0$, namely, for $j=0$:
%
%
%\begin{equation}
%        \begin{split}
%			\beta_2(\mu_I)&=\dfrac{1}{2}F_\pi^2\left[M_\pi^2-\left(1-a_1\right)\mu_I^2\right]+\dfrac{1}{2}\left[\dfrac{M^2}{2}(\hat{q}_2^r+3\hat{q}_3^r)\mu_I^2+2\hat{q}_6^r\mu_I^4\right]\\
%			&+\dfrac{a_1}{64\pi^2}\left[\mu_I^4-\dfrac{M^2}{2}(1+8\bar{l}_4)\mu_I^2+2\left(5M^2-\mu_I^2\right)\mu_I^2\log\left(\dfrac{\mu^2}{M^2}\right)\right]\\
%			&+\dfrac{a_1^2}{32\pi^2}\mu_I^4\left[\dfrac{\mu_I^2}{2M^2}-\dfrac{9}{2}+3\log\left(\dfrac{\mu^2}{M^2}\right)\right]\\
%			&-\dfrac{F^2}{2}\left[M^2\cos(\alpha_0^{LO})-(1-a_1)\mu_I^2\cos(2\alpha_0^{LO})\right]+\mathcal{O}\left(\dfrac{1}{F_\pi^2}\right)+\mathcal{O}(a_1^3)
%		\end{split}
%\label{BECnlo}
%\end{equation}

% Note that, because of the renormalization conditions \eqref{qren}-\eqref{qrencoeff}, we have to keep up to $a_1^2$ terms in order to ensure that the resulting expressions are finite and scale-independent. 

From the previous expression, we see that only $\hat{q}^r_{2,3,6}$ and $a_1$ modify the critical BEC value.   We show in Table \ref{becnatvalues} the value of $\mu_c$ expected within the range of natural
values for those constants, defined within a typical ChPT uncertainty range \cite{Knecht:1997jw,Meissner:1997fa} as 

\begin{equation}
\vert \hat q_i^r \vert \leq \frac{1}{16\pi^2}
\label{natural}
\end{equation}
where we have set the scale as $\mu=M_\rho\simeq$ 770 MeV and we have taken $a_1=-0.019$, the central value of the LO fit 3  to lattice results in Section \ref{sec:LOfits}.  
In addition, we have highlighted in the table the values for which the condition \eqref{extracond} holds. As explained, $\hat q_i^r$ values not satisfying  that condition are not acceptable, since they give rise to a nonzero isospin density below the critical value. 

Note that, as happened at LO, our results support $\mu_c<M_\pi$. In turn, we remark that the above analysis would allow in principle to fix $\mu_c=M_\pi$, as in previous ChPT studies \cite{Adhikari:2019mdk,Adhikari:2020ufo}, by  imposing an additional constraint relating $\hat{q}^r_{2,3,6}$ and $a_1$ from \eqref{beta2}, namely

\begin{equation}
\mu_c^{NLO}=M_\pi \Rightarrow a_1F_\pi^2+\dfrac{M_\pi^2}{2}\left[\hat{q}_2^r+3\hat{q}_3^r+4\hat{q}_6^r 
%-\dfrac{F^2 \left(\mu_I^4-M^4\right)}{2\mu_I^2}
%-F^2 (M_\pi^2-M^2)
+\dfrac{\bar l_3}{8\pi^2}  \right] 
		=0+\mathcal{O}\left(\frac{1}{F^2}\right)
		\label{masscond}
\end{equation} 
where $\bar l_3$ arises from the $M_\pi^2-M^2$ difference \cite{Gasser:1983yg} and we have replaced $M$ by $M_\pi$ in the NLO when the difference is of higher order. 

Nevertheless,  as explained in previous sections, lattice results are actually compatible with $\mu_c<M_\pi$ and in principle there is not  a  physical reason to impose the above constraint, unlike those coming from the vanishing of the isospin density and discussed in section \ref{sec:const},  so we will allow the $\hat q_i^r$ to fluctuate within natural values  without imposing \eqref{masscond} when evaluating the different observables in the following section.

%We only get a critical value slightly higher than $M_\pi$ for $\hat{q}_{3,6}^r=\frac{1}{16\pi^2}$ and $\hat{q}_{2}^r=\pm\frac{1}{16\pi^2}$. 

%In addition, we have checked that the condition $ \beta_2(\mu_I)>0$ is fulfilled  for $\mu_I<\mu_c$ for the values of $\hat q_i^r$ in Table \ref{becnatvalues}. 
%\anote{Esta frase queda un poco rara, habr\'ia que decir algo entonces tambi\'en sobre el valor l\'imite de $\mu_I$ en ese rango de LEC, es decir, hasta que valor de $\mu_I$ existe m\'inimo. A\'un no me queda claro, por definici\'on $\beta_2$ no deber\'ia cambiar de signo por debajo de $\mu_c$ porque entonces $\mu_c$ cambiar\'ia. }

\begin{table}[h!]
\begin{center}
\begin{tabular}{|c|c|c||c|c|}
\cline{2-5}
\multicolumn{1}{c|}{}& \multicolumn{2}{c||}{$\hat{q}_6^r=\frac{1}{16\pi^2}$}&
\multicolumn{2}{c|}{$\hat{q}_6^r=-\frac{1}{16\pi^2}$}\\
\cline{2-5}
\multicolumn{1}{c|}{$\mu_c^{NLO}/M_\pi$}& \multicolumn{1}{c|}{$\hat{q}_3^r=\frac{1}{16\pi^2}$} & \multicolumn{1}{c||}{$\hat{q}_3^r=-\frac{1}{16\pi^2}$} &\multicolumn{1}{c|}{$\hat{q}_3^r=\frac{1}{16\pi^2}$} &\multicolumn{1}{c|}{$\hat{q}_3^r=-\frac{1}{16\pi^2}$}\\
\hline
$\hat{q}_2^r=\frac{1}{16\pi^2}$ & 1.007& \textbf{0.935}& \textbf{0.916}& \textbf{0.803}\\
\hline
$\hat{q}_2^r=-\frac{1}{16\pi^2}$ & 1.005& \textbf{0.873}& \textbf{0.874} & \textbf{0.773}\\
\hline
\end{tabular}
\end{center}
\caption{Critical value of the BEC transition for natural values of $\hat{q}_2^r$, $\hat{q}_3^r$ and $\hat{q}_6^r$. The highlighted values are those fulfilling the condition \eqref{extracond}.}
\label{becnatvalues}
\end{table}

\subsubsection{NLO Results for different observables}

We will consider now the  NLO evolution of chiral observables for nonzero $\mu_I$, in particular regarding the role of the $\hat{q}_i^r$ LEC and the comparison with lattice analyses. From the results in the previous sections, we see that the NLO energy density depends on seven independent new LECs, namely $a_1,\hat{q}_{1-6}^r$ whose numerical values will then influence the $\mu_I$ dependence of the observables.  Note  that all the observables considered depend on $\hat{q}_{1-6}^r$ since, in addition to the explicit $\hat q_i^r$ dependence one must consider that in $\alpha_0^{NLO}$. We will represent our results for the range of natural values \eqref{natural}  at the scale $\mu=M_\rho$ and setting $a_1=-0.019$ (the mean value obtained in Fit 3 in section \ref{sec:LOfits}). Thus, we have  $3^6$ points for each $\mu_I$, corresponding to the three values $0, \pm1/(16\pi^2)$. In doing so, we  discard those $\hat q_i^r$ violating the condition \eqref{extracond} and we calculate the mean square error of the results which provides a dispersion estimation. 
Note that, as commented in section \ref{sec:lin},  due to the linear term, the minimum could disappear above a given $\mu_I$. Thus, the uncertainty bands in the following figures for a given $\mu_I>\mu_c$ correspond to those $\hat q_i^r$ combinations for which the minimum exists. The effect of the linear term will be actually showed separately in the figures in order to calibrate better its effect.

\begin{figure}[h!]
\begin{center}
  \includegraphics[width=0.48\textwidth]{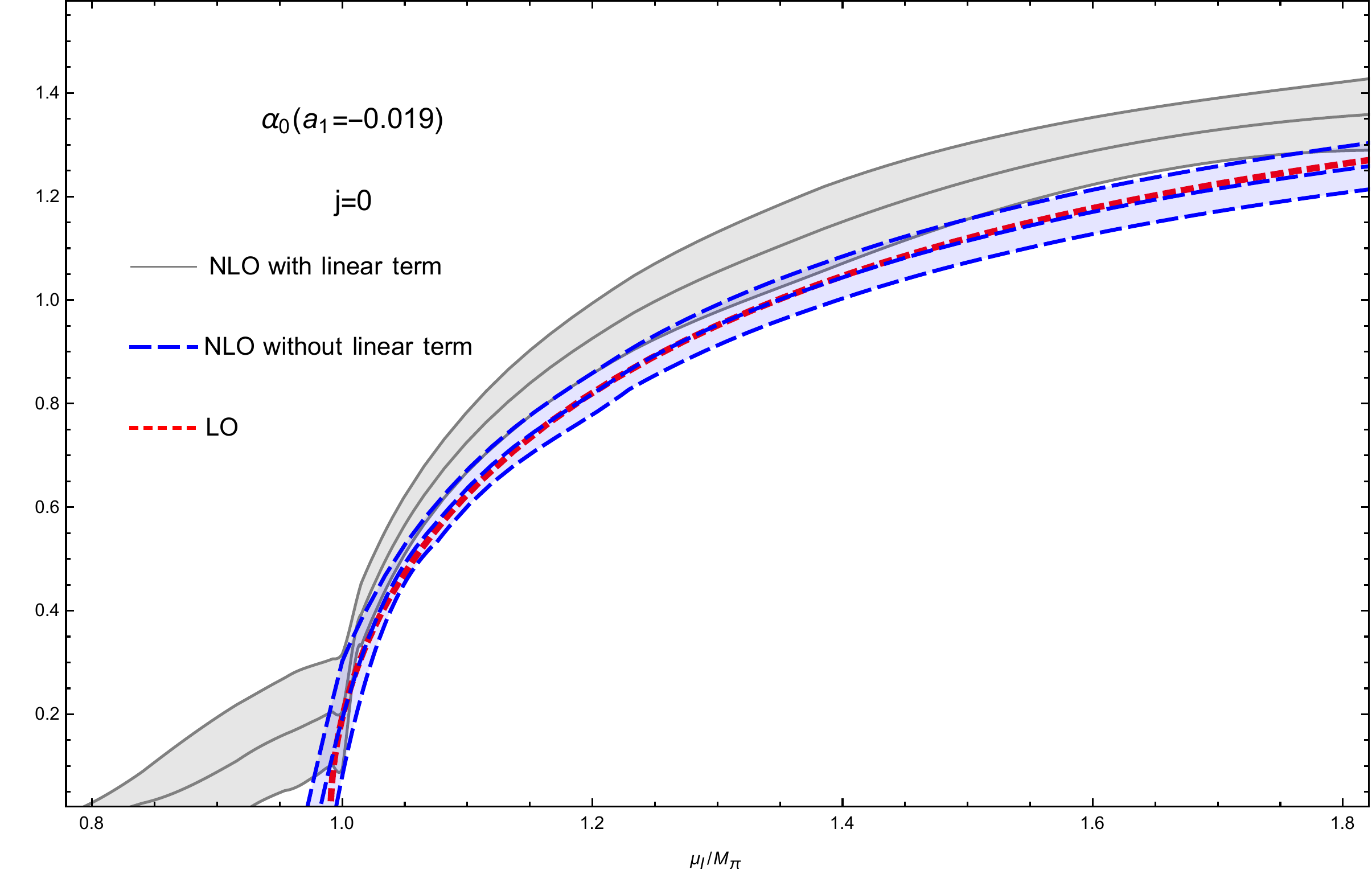}
  \includegraphics[width=0.48\textwidth]{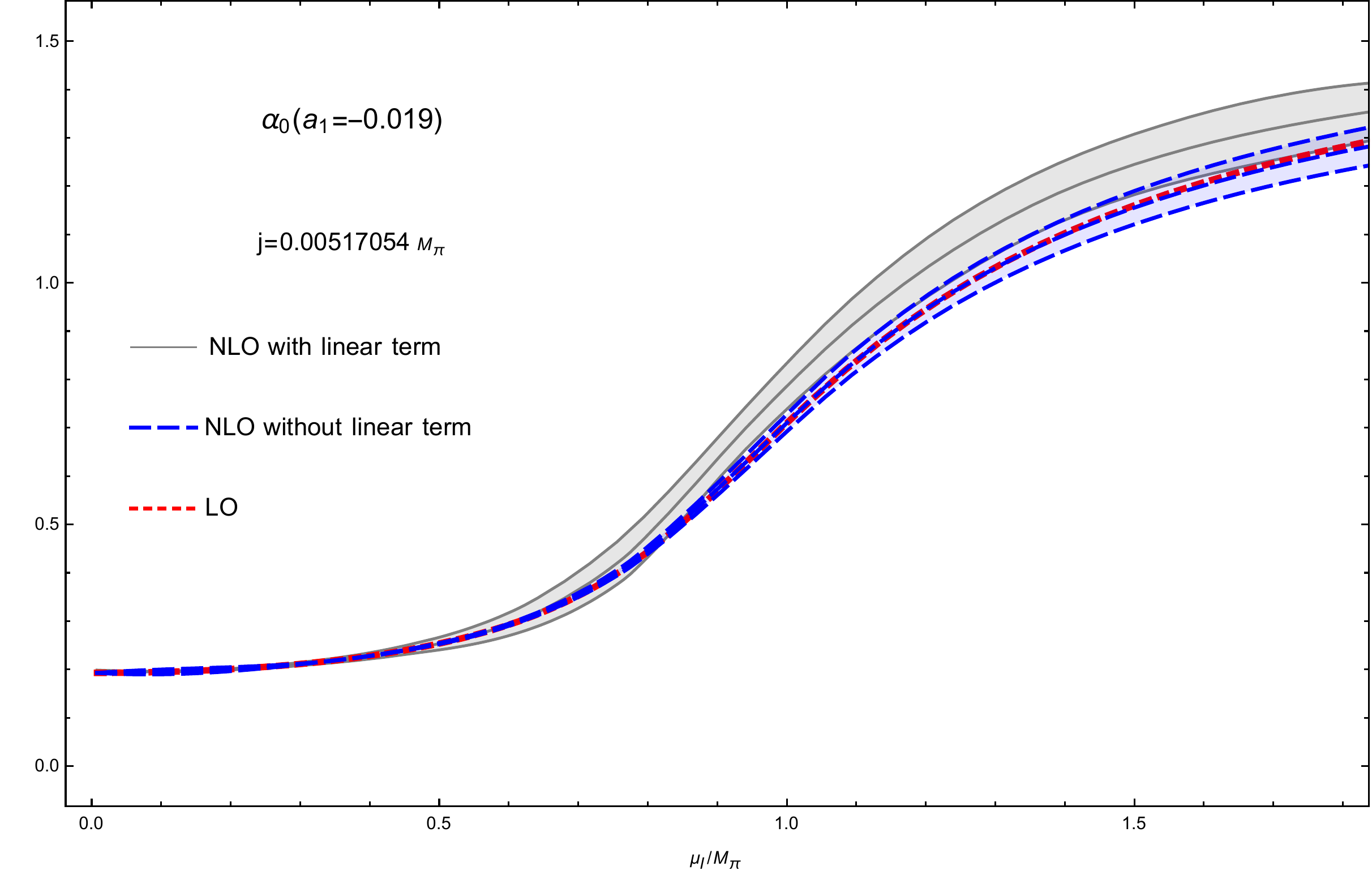}
  \end{center}
  \caption{Results for the minimizing angle $\alpha_0$ at NLO for finite source $j=0$ (left) and $j=0.005170554 M_\pi$  (right), with $a_1=-0.019$ and natural values for $\hat{q}_{1-6}^r$.}
  \label{alpha0qs}
\end{figure}

First, in Fig.. \ref{alpha0qs}, we show $\alpha_0^{NLO}$, i.e., the angle minimizing the energy density including NLO corrections.
We see that  including all the corrections discussed here implies sizable deviations above $\mu_c$ with respect to the LO, larger than in previous analyses  \cite{Adhikari:2019mdk,Adhikari:2020ufo}. Note in particular the tail below  $\mu_I=M_\pi$ coming from the reduction in the numerical value of $\mu_c^{NLO}$ as Table \ref{becnatvalues} shows.

%The minimum of this quantity is independent of $\hat{q}_7^r$ because the term multiplying this constant does not depend on $\alpha$. 
%In section \ref{sec:lo}. we have analysed the value of the $a_1$ constant. 

%On the other hand, from  the terms dependent on the quark mass in \eqref{energy4qsa1ren}, we see that the  LEC $a_1$, $\hat{q}_{2,3}^r$ contribute to the quark condensate dependence on $\mu_I$.  Likewise, the pion condensate depends on $a_1$, $\hat{q}_{4,5}^r$ since those are the only terms depending explicitly on the pionic source $j$. In addition, we must also account for the dependence of those condensates on $\hat{q}_{1-6}^r$ through $\alpha^{NLO}$, as mentioned above. Altogether, both observables depend on $\hat q^r_{1-6}$. 

\begin{figure}[h]
  \includegraphics[width=0.48\textwidth]{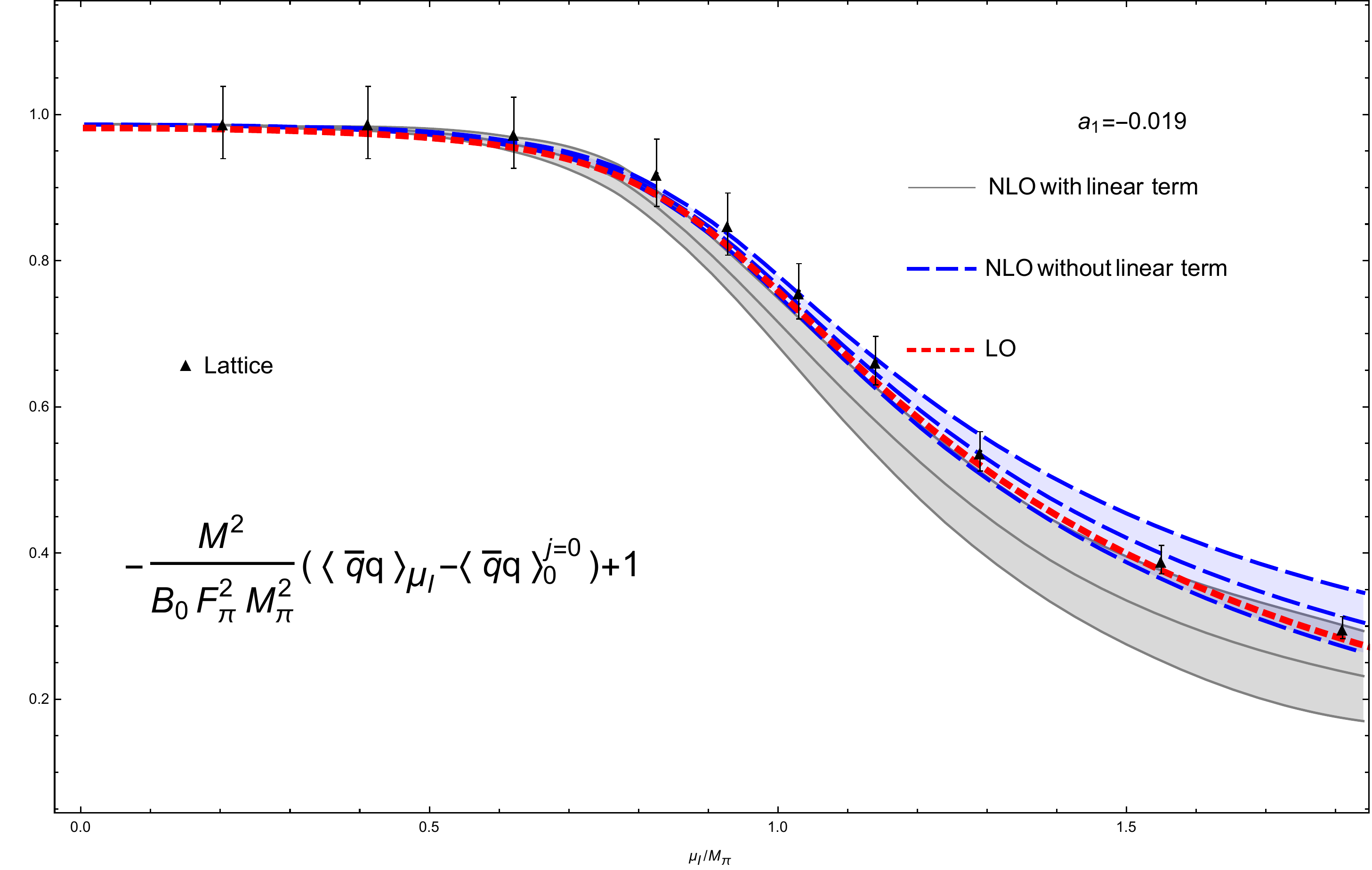}
  \includegraphics[width=0.48\textwidth]{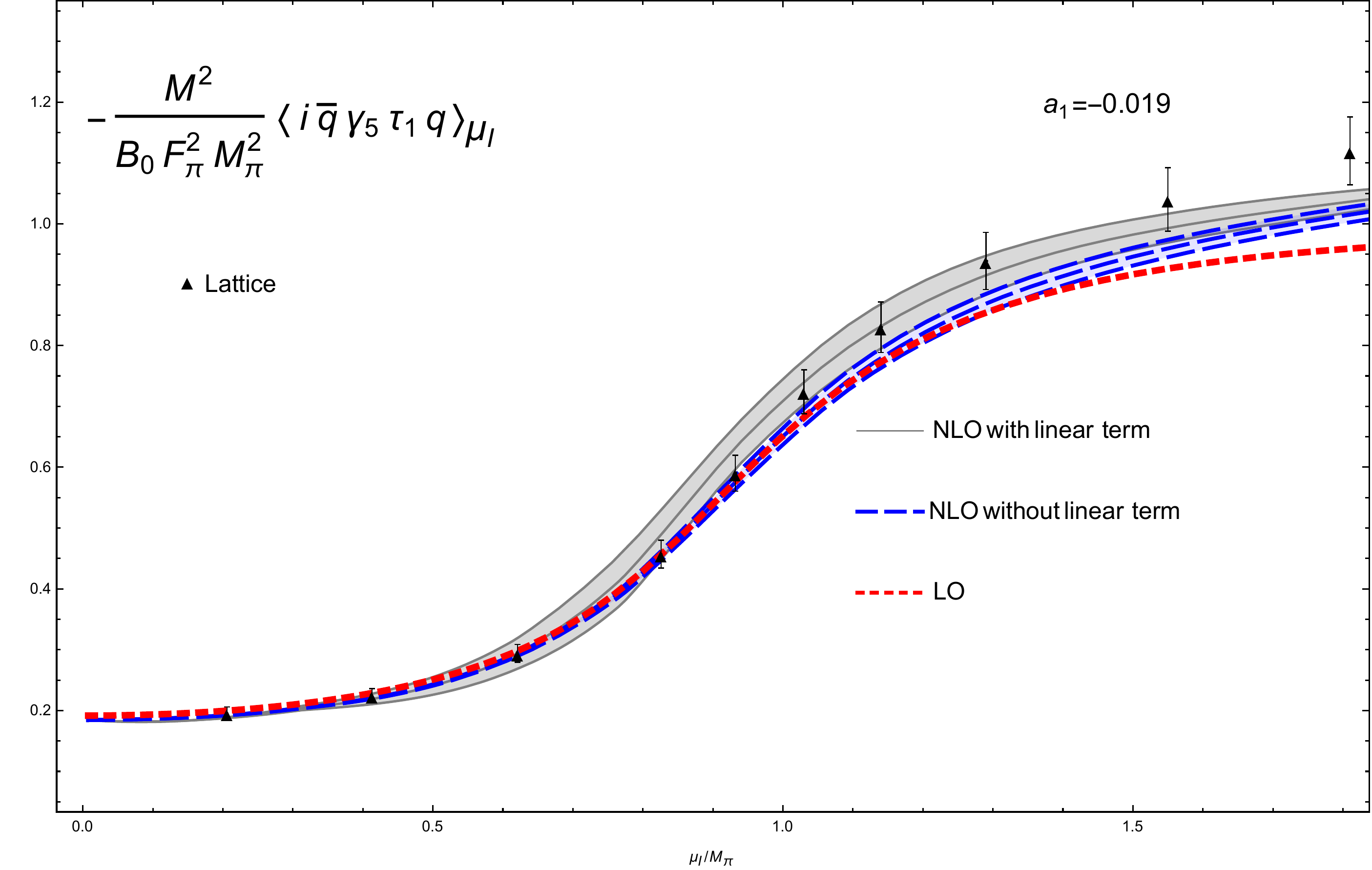}
  \caption{Results for the quark condensate deviation (left) and pion condensate deviation (right) as a function of $\mu_I/M_{\pi}$ at NLO for finite source $j=0.005170554 M_\pi$, $a_1=-0.019$ and natural values for $\hat{q}_i^r$. The lattice data are from \cite{Adhikari:2020ufo}.}
  \label{qqnlowithlinqs}
\end{figure}

 We plot in Fig. \ref{qqnlowithlinqs} the quark and pion condensate deviations (as defined in \cite{Adhikari:2020ufo}) at NLO for natural values of $\hat{q}_i^r$ with and without linear term, comparing  with lattice results.   The NLO corrections are again significant and remain within the uncertainties of lattice points, taking into account that we are not performing a NLO fit so there would be still room for improvement.  The NLO corrections actually improve over the LO fit for the pion condensate for high $\mu_I$ values, whereas in the case of the quark condensate, with the inclusion of the linear term, the theoretical curve seems to depart from the lattice points with respect to the LO.

\begin{figure}[h!]
  \includegraphics[width=0.8\textwidth]{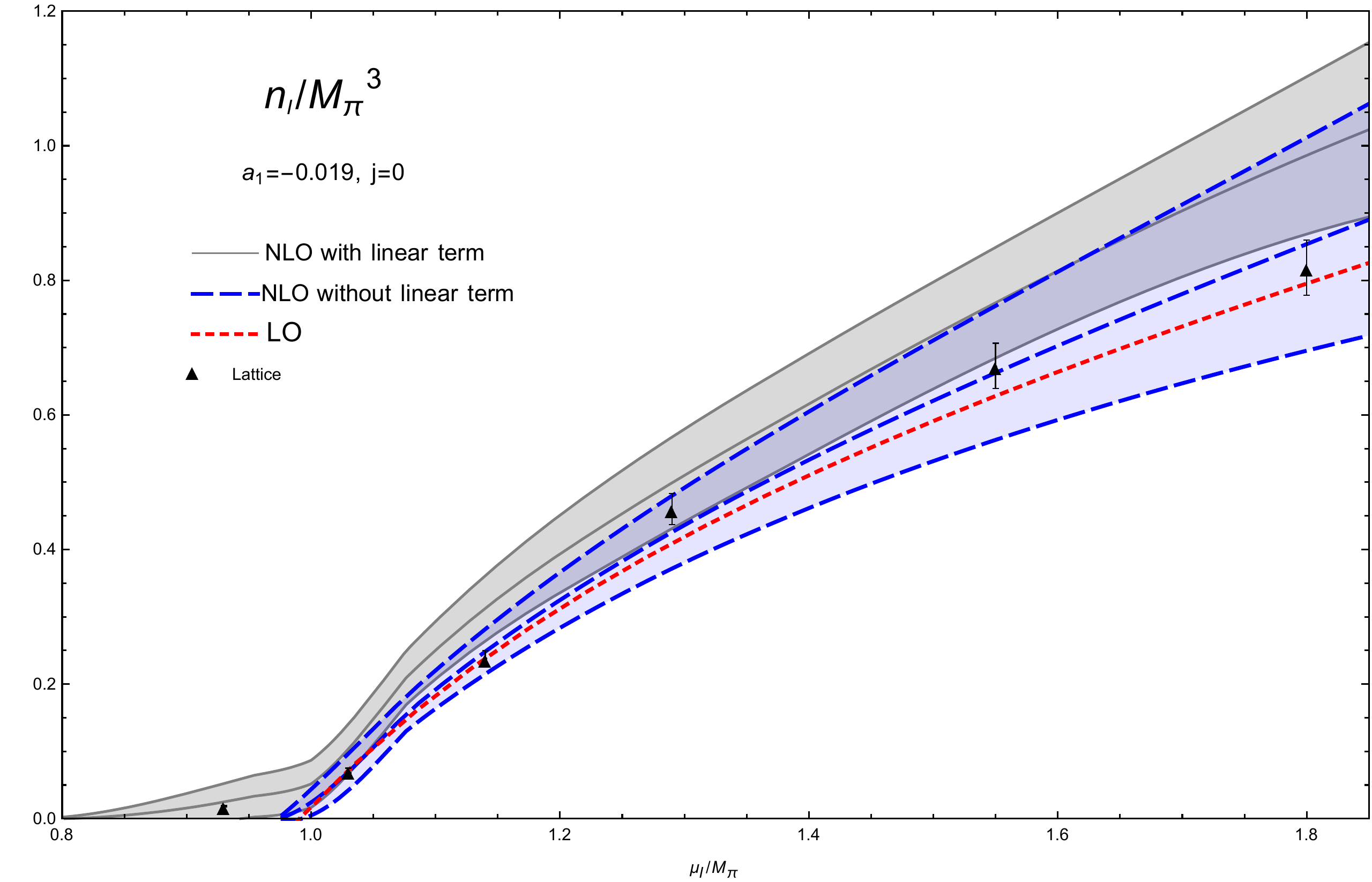}
  \caption{Normalized isospin density as a function of $\mu_I/M_{\pi}$ at NLO with and without the  linear term. Lattice points are taken from \cite{Brandt:2018bwq}.}
  \label{nInlowithlinqs}
\end{figure}

Finally, we study the isospin density at NLO for $j=0$. The result, including the uncertainty bands of $\hat{q}_i^r$ natural values, are showed in  Fig. \ref{nInlowithlinqs}. As for previous observables, the NLO with the new terms considered in this work provides significant deviations from the LO as $\mu_I$ increases, still accommodating the lattice results.

As explained above, for the NLO analysis we have fixed $a_1$ to the LO fit and consider the new $\hat{q}_i^r$ within natural values. In view of the results  in Figs. \ref{qqnlowithlinqs} and \ref{nInlowithlinqs}, we would surely have obtained a much better description of lattice results by keeping $a_1$ and the $\hat{q}_i^r$ as fit parameters with the NLO curves, but such precision analysis is outside the scope of this work. 
%Besides, the fact that the minimum disappears for certain $\hat{q}_i^r$ combinations and the constraint  \eqref{extracond} would make the analysis more complicated. 

%Actually, from the previous curves and the analysis in section \ref{sec:lo}, we see that the effect introduced by linear term shows a certain similarity to the effect of having a negative $a_1$  and therefore  a higher negative value of $a_1$ would improve the results. 

\section{Conclusions}

In the present work, we have analyzed the most general effective chiral lagrangian for nonzero isospin chemical potential for two light flavours up to fourth order. We have followed the technique of external sources including spurion fields, which allows to account for all possible operators respecting the symmetry breaking problem at hand. We have analyzed in detail the effect of new terms with respect to previous analyses.

In particular, at leading order, two new-independent terms in the ${\cal L}_2$ lagrangian have to be considered, whose corresponding low-energy constants $a_1$, $a_2$ are related from the constraint of vanishing isospin density below the critical BEC value $\mu_c$. The constant $a_1$ generates a shift in $\mu_c$ with respect to the pion mass at that order. To estimate the preferred value of $a_1$, we have performed several fits to lattice results for the quark and pion condensates (for nonzero pionic source $j$) as well as for the isospin density (for $j=0$). When comparing to the $a_1=0$ results, a small nonzero negative value for $a_1$ is favored, improving the description of lattice results, specially for the isospin density.

To NLO, the fourth order lagrangian ${\cal L}_4$ contains seven new terms multiplied by  new low-energy constant $\hat{q}_i$, which we have consistently renormalized to absorb the divergences coming from loops with vertices of the new ${\cal L}_2$ terms. Apart from those terms, we have shown that to NLO one has to consider an additional contribution coming from a term in ${\cal L}_2$ linear in the pion fields. That term does not contribute to LO but it does to NLO due to the corrections in the angle minimizing the energy density. The  effect of the linear term is  qualitatively important, since it eventually makes the minimum of the energy density disappear, which sets a natural limit of validity for the ChPT framework, consistently with lattice analyses. 

Imposing that the isospin density vanishes below $\mu_c$ to NLO gives rise to additional constraints, involving now the $a_1,a_2,\hat q_i^r$ LEC.  Those constraints allow on the one hand to understand the small value for $a_1$ obtained in the lattice fits and on the other hand, to eliminate one of the seven new LEC. In addition, the critical BEC value to NLO must remain below the LO one, which restricts further the admisible values for the $\hat q_i^r$. A further constraint could be obtained by demanding $\mu_c=M_\pi$.  We have estimated the effect of the new LEC and the linear term to NLO by keeping the $\hat q_i^r$ within natural values. The results for the different observables show again consistency with the lattice points, leaving room for improvement with respect to the LO. 

In summary, our present analysis establishes systematically the most general way to describe low-energy QCD at nonzero isospin density, consistently with lattice results and complementing previous theoretical analyses. We believe that this work will be useful towards a better understanding of the QCD phase diagram and we leave for future works its extension to three flavours and finite temperature.

% {\it Acknowledgments}---
 
 \begin{acknowledgments}
 Work partially supported by research contract  PID2019-106080GB-C21 (spanish ``Ministerio de Ciencia e Innovaci\'on"), and the European Union Horizon 2020 research and innovation program under grant agreement No 824093. A. V-R acknowledges support from a fellowship of the UCM predoctoral program.
 \end{acknowledgments}

\end{document}